\documentclass[lettersize,journal]{IEEEtran}
\usepackage{amsmath,amsfonts}

\usepackage{array}

\usepackage{textcomp}
\usepackage{stfloats}

\usepackage{verbatim}
\usepackage{graphicx}
\usepackage{cite}
\usepackage[ruled]{algorithm2e}
\usepackage{algorithmicx}
\usepackage{mathptmx}
\usepackage{caption}
\usepackage{subcaption}
\usepackage{url}
\usepackage{hyperref}
\usepackage[capitalize]{cleveref}
\usepackage{xcolor}
\begin{document}

\title{A Coverage Control-based Idle Vehicle Rebalancing Approach for Autonomous Mobility-on-Demand Systems}

\author{Pengbo Zhu, Isik Ilber Sirmatel, Giancarlo Ferrari-Trecate, Nikolas Geroliminis
  % <-this % stops a space
\thanks{Pengbo Zhu and Nikolas Geroliminis are with the Urban Transport Systems Laboratory, \'Ecole Polytechnique F\'ed\'erale de Lausanne, 1015 Lausanne, Switzerland (email:  {\tt\small pengbo.zhu@epfl.ch,  nikolas.geroliminis@epfl.ch}).

Isik Ilber Sirmatel is with the Control Section, Department of Electrical and Electronics Engineering, Faculty of Engineering, Trakya University, 22030 Edirne, T{\"{u}}rkiye (email: {\tt\small iilbersirmatel@trakya.edu.tr}).

Giancarlo Ferrari-Trecate is with the Dependable Control and Decision Group, \'Ecole Polytechnique F\'ed\'erale de Lausanne, 1015 Lausanne, Switzerland (email: {\tt\small giancarlo.ferraritrecate@epfl.ch}).}}% <-this % stops a space
%\thanks{Manuscript received April 19, 2021; revised August 16, 2021.}}

\markboth{}%
{ZHU \MakeLowercase{\textit{et al.}}: A Coverage Control-based Idle Vehicle Rebalancing Approach for Autonomous Mobility-on-Demand Systems}


\maketitle

\begin{abstract}
As an emerging mode of urban transportation, Autonomous Mobility-on-Demand (AMoD) systems show the potential in improving mobility in cities through timely and door-to-door services. However, the spatiotemporal imbalances between mobility demand and supply may lead to inefficiencies and a low quality of service. Vehicle rebalancing (i.e., dispatching idle vehicles to high-demand areas), is a potential solution for efficient AMoD fleet management. In this paper, we formulate the vehicle rebalancing problem as a coverage control problem for the deployment of a fleet of mobile agents for AMoD operation in urban areas. Performance is demonstrated via microscopic simulations representing a large urban road network of Shenzhen, China. Results reveal the potential of the proposed method in improving service rates and decreasing passenger waiting times.
\end{abstract}


\section{Introduction}

\IEEEPARstart{T}{he} continuous expansion of modern urban space and worldwide increasing population density in mega-cities lead to rapidly growing mobility demand. Addressing this problem exclusively via private vehicles is not sustainable owing to the increase in congestion and emissions. Moreover, public transportation alone cannot satisfy the need for timely, comfortable, and door-to-door mobility services. New trends in shared transportation are seen to have the potential to change the landscape and understanding of mobility. Mobility-on-Demand (MoD) systems (such as Uber, Lyft, and Didi) are promising solutions for providing passengers with efficient and fast service by deploying a group of coordinated vehicles serving ride requests within city areas. Using autonomous vehicles as MoD fleet (Autonomous Mobility-on-Demand, AMoD) is expected to improve upon these solutions \cite{Zardini2022}. 
  
The rapid adoption of MoD systems creates great challenges for city mobility operators and policymakers as there is limited information on the decision trends, safety implications, impact on vehicle ownership, and the effect of travel patterns on congestion. Furthermore, Transportation Network Companies (TNCs) are trying to attract more demand from other modes (e.g., taxis and buses) by offering shorter waiting times via using larger fleets, which can cause significant delays in the network due to a large number of circulating vehicles without passengers \cite{Caio2021}. Thus, efficient MoD fleet management is an important direction for research, as it can enable TNCs to achieve the same service quality with smaller fleets, leading ultimately to less congested networks and thus shorter delays for all participants of urban mobility. 
% Although there are many methods proposed for MoD systems (e.g., vehicle-sharing \cite{AlonsoMora2017A}, routing \cite{AlonsoMora2017B}, dynamic fare and pricing \cite{NOURINEJAD2020}, fleet sizing \cite{Alex2019}), relatively few studies have investigated 
% A considerable literature has grown up around MoD systems, 
Many studies have been investigated focusing on different aspects of MoD systems,
% Many aspects relevant to MoD systems have been 
for example, vehicle-sharing \cite{AlonsoMora2017A}, routing \cite{AlonsoMora2017B}, dynamic fare and pricing \cite{NOURINEJAD2020}, and fleet sizing \cite{Alex2019}.
Moreover, in recent years, there has been an increasing interest in the spatial rebalancing problem of MoD fleets.

Asymmetry between origin and destination distributions of trips and non-uniform passenger demand for rides in different districts create discrepancies between actual and desirable spatial distributions of MoD fleets. For example, in the morning, people will typically travel from their houses to work, and in the evening, the trips will mostly be in the opposite direction. Obviously, more trips will begin or end around central business districts while relatively few will be in rural areas. Thus, proactively relocating idle vehicles to high-demand areas in real-time (i.e., vehicle rebalancing) can have significant improvements on MoD system performance \cite{Alex2018, Marczuk2016}. Nowadays, most TNCs use applications on mobile phones to receive requests from passengers and assign these orders to drivers \cite{Caio2021}. Since a passenger will not wait indefinitely to be picked up, and drivers might refuse to answer a low-profit request, the imbalance of supply and demand can lead to long passenger waiting times and high cancellation rates, harming service quality. In MoD systems, there is a clear advantage in deploying vehicle rebalancing algorithms to achieve efficient operation by sending idle vehicles to districts with current or future high demand. Most literature and applications for rebalancing mainly focus on car-sharing and bike-sharing systems with much smaller fleet sizes \cite{ILLGEN2019, REPOUX2019}. Bike-sharing systems are relatively easy to operate due to the size of bikes allowing a single mini-van to relocate a large number of bikes simultaneously \cite{Forma2015, Fagnih2017}. The management of car-sharing systems has been addressed at the strategic, tactical, and operational levels. For comprehensive reviews on these three planning levels, see \cite{Laporte2015}. The strategic and tactical decision levels relate primarily to station location \cite{CORREIA2012,Burak2015}, fleet sizing \cite{Cepolina2013}, and staff sizing \cite{Kek2009}. At the operational level, most of the emphasis has been put on devising methods to redistribute vehicles in the system. 
However, the relocating of sharing bikes or cars mostly happens once per day at midnight, thus such methods are not suited for the real-time rebalancing of MoD systems.

Methods for MoD system management have been receiving increasing attention in the literature. In \cite{Pavone2010}, by partitioning city districts into small areas (denoted as stations), a high-load case is presented to deploy a number of vehicles at each station for an AMoD system. Stations are modeled as vertices of a graph and a rebalancing policy is applied between vertices. A fluid discrete-time approximation model is introduced and an optimization approach is presented for steering the system to equilibrium \cite{Pavone2011, Pavone2012}. Model predictive control (MPC) approaches are employed to determine the optimal rebalancing policy in \cite{Andrea2019, Mohsen2018}, while \cite{Zhang2016} provides proof of stability within an MPC framework. \cite{Miao2016} utilizes both historical and real-time data to build a prediction model. In \cite{AlonsoMora2017B}, the rebalancing problem is described as an integer linear programming problem to match vehicles with passengers and balance the fleet. For shared MoD systems, a model-free reinforcement learning scheme is proposed to offset the imbalance in \cite{Jian2017}, while it has the risk of dispatching more vehicles than needed to an area. 
In most of the previous literature, a centralized controller is considered to operate the fleet. Moreover, as most works are based on region/station level, partitioning the city area into a limited number of virtual stations is required as a preprocessing step \cite{CHEN2021}.

The application of advanced control methodologies is receiving attention for efficient management of traffic networks (see, e.g., \cite{Pasquale2021}). For multi-agent systems, motion coordination problems have been investigated, which automatically distribute a fleet of mobile agents (such as autonomous vehicles or mobile robots) to carry out certain tasks in a bounded environment (e.g., rendezvous maneuver \cite{Mancini2020}, task allocation, pattern formation \cite{Sonia2011}). 
Coverage control methods have been investigated to find the spatial configuration of agents that optimize a prescribed cost\cite{Vishaal2022}. 
By assuming each mobile agent has a uniform circular sensing footprint, \cite{Sotiris2016, Sotiris2018A, Sotiris2018B} focus on maximizing area coverage over the region of interest.
The purpose of  \cite{Jorge2004} is to minimize the sum of the demand-weighted distance between the agent and all points in its covered area. In \cite{CARRON2020}, nonlinear dynamics of mobile agents were considered by using model predictive control schemes. Additionally, the coverage control problem has been expanded to consider constraints such as collision avoidance\cite{Hussein2007}, energy consumption\cite{Ru2012}, and limited communication range\cite{SONG2018}. The majority of previous works are based on dynamic Voronoi partition, whose Voronoi generators are the current positions of agents. In \cite{Jorge2004}, Lloyd's algorithm \cite{Lloyd1982} is used as a classic approach to drive agents to converge to a Centroidal Voronoi Configuration (CVC), with the center of mass of each Voronoi cell (i.e., centroid) coinciding with its generator \cite{Qiang1999}. 
The aforementioned works focus only on agents moving in a continuous Euclidean space. By considering a more practical setting, \cite{Yun2012} and \cite{Durham2012} use graph Voronoi partition \cite{Erwig2000} to represent a complex non-convex spatial environment. 

There is great potential in applying coverage control to AMoD systems,
as it can be operated in a distributed way with each agent computing its own control action, specifying better scalability properties compared to centralized schemes \cite{Fatemeh2016,chaozhe2020}. Without pre-partitioning of the city area, the rebalancing position control can be made more precise via node-level (instead of region/station level) rebalancing.
Furthermore, coverage control can run in real-time which is suitable for continuously allocating idle vehicles according to current idle fleet size and demand conditions.

Building upon our earlier conference work \cite{Pengbo2022}, 
% this paper extends the applied control literature in the following aspects:
this paper provides the following contributions:

1) To the best of our knowledge, it describes the first attempt at applying coverage control \cite{Jorge2004} to a real-time vehicle fleet rebalancing problem. The proposed method enables efficient AMoD system operation via dispatching idle vehicles towards high-demand areas, even the number of active vehicles involved in coverage is continuously changing due to passenger drop-offs and pick-ups. Furthermore, we validate the performance of the proposed method across various imbalance levels in trip origin and destination distribution, as well as different controller sampling times.

2) We extend the coverage control method to graphs (via \cite{Durham2012}). This is important for the practical implementation of the proposed algorithm in urban road networks that invariably have a graph structure (while the original coverage control method involves a continuous space). The performance of these two methods is compared from different perspectives, e.g., computational complexity, implementation requirements, and so on.

3) We propose two kinds of upper-level controllers for choosing a subset of the idle vehicle fleet to hold position. This is required for reducing excessive rebalancing travel distance, e.g., back-and-forth motion, which might be caused by exactly following the coverage control actions in continuously changing traffic conditions. 
Using such an approach can mitigate the adverse effects of unnecessary traffic congestion and fuel consumption.

The remainder of the paper is organized as follows: Mathematical preliminaries of the coverage control method are introduced in \cref{Preliminary}. Details of applying coverage control to vehicle fleet rebalancing are provided in \cref{problemformulation}. In \cref{Experiments}, simulation results, obtained using an MoD system simulator and synthetic scenarios with different imbalance levels, verify the proposed methods can answer more trip requests with less waiting time. Then, \cref{Extension} considers the application of coverage control on the graph instead of on the continuous space. In addition, two controllers for determining active idle fleet size are introduced in \cref{fleetsizing}, and some results are provided showcasing their operations. Finally, \cref{conclusion} provides conclusions and potential directions for future work.

\section{Mathematical Preliminaries}\label{Preliminary}
In this section, we state the mathematical preliminaries and definitions of Voronoi partition and centroidal Voronoi configuration \cite{Qiang1999}, and recall the control algorithm `move-toward-the-centroid of each Voronoi cell' proposed in \cite{Jorge2004}. 
\subsection{Voronoi Partition}

\begin{figure}[ht]
     \centering
         \includegraphics[width=0.40\textwidth]{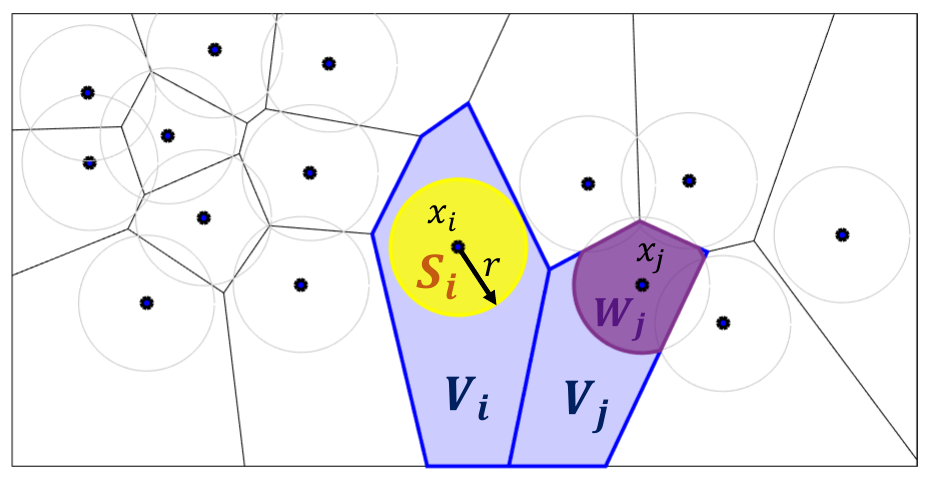}
         \caption{Illustration of Voronoi partition.}
         \label{fig:final}
         \vspace{-0.5cm}
     \end{figure}

Consider a bounded convex set $\Omega \subset \Re^2$. A number of $n$ mobile agents are deployed, whose positions are $X = \{x_1, x_2, ..., x_n\}, x_i \in \Omega$ and their movements are constrained inside $\Omega$. A naturally complete partition of $\Omega$ is the Voronoi partition, which gives a tessellation related to the responsibility area for an agent (Voronoi cell). Each cell consists of all points \(q\) in \(\Omega\) that are closer to agent \(x_i\) than to any other agent. It is defined as 
% Consider the urban area as a bounded convex space $\Omega \subset \Re^2$. A number of $n$ idle vehicle are deployed whose positions are $X = \{x_1, x_2, ..., x_n\}, x_i \subset \Omega$ and their movement are constrained inside $\Omega$.  A naturally complete partition of $\Omega$ is Voronoi partition, which can give a tessellation that the responsibility area for an agent (Voronoi cell) is the area in $\Omega$ including all points that are closer to it than any other agent. It is defined as 
\begin{multline}
        V_i(x_i) = \{q \in \Omega: \Vert x_i - q \Vert \leq \Vert x_j - q \Vert, \forall j \neq i \}.
\end{multline}
where $\Vert \cdot \Vert$ is the Euclidean norm and  $x_i$ is called the generator/seed of the Voronoi cell $V_i(x_i)$ (as an example, $V_i(x_i)$ and $V_j(x_j)$ are shown as blue polygons in \cref{fig:final}). A Voronoi partition $V = \{V_1, V_2, \dots, V_n\}$ has the property that
\begin{equation}
        \bigcup_{i=1}^n V_i = \Omega, 
    \qquad int\{V_i\}\cap int\{V_j\}=\emptyset, \quad \forall i \neq j,
\end{equation}
where $int\{V_i\}$ denotes the interior of $V_i$. 
     
\subsection{Centroidal Voronoi Configuration}
A continuous integrable function $\phi: \Omega \rightarrow \Re$ is termed a distribution density function and used for representing a measure of information or the probability of occurrence of an event at each point in $\Omega$. For each Voronoi cell $V_i, i \in \{1,2, \dots, n\}$, the center of mass (i.e., centroid) is defined with respect to the density function $\phi$ as
\begin{equation}\label{centroidV}
     C(V_i) = \frac{{ \int_{q \in V_i}} q\phi(q)dq}{{\int_{q \in V_i}} \phi(q)dq},
\end{equation}

If the position of each agent $x_i, i \in {1, 2, \dots, n}$ is simultaneously located at its centroid, 
% this critical partition $V$ and spatial configuration $X$ 
the corresponding pair $(V, X)$ is called \textit{centroidal Voronoi configuration} \cite{Qiang1999}.
 
\subsection{The Centroidal Coverage Control Law}
The coverage objective function $H$: $\Omega \rightarrow \Re$ can be formulated as follows,
\begin{equation}\label{coverageobjV}
    H(X,V) = \sum_{i= 1}^n \int_{q \in V_i }f(\Vert x_i - q \Vert)\phi(q)dq,
\end{equation}
where $f: [0,\infty)\rightarrow \Re$ is a nondecreasing differentiable cost which describes how the coverage performance degrades with the distance $\Vert x_i - q \Vert$ between an agent and a given point $q$. As an example, one can set $f(x) = x^2$, which is generally used in facility location problems\cite{Jorge2004}. 
% For instance, when choosing the location of hospitals in order to minimize the total transportation cost satisfying the demands of patients from different communities.

% Considering simple dynamics for the agent motion as follows
We assume that agent $i\in\{1,2,\dots,n\}$ obeys the integrator dynamics,
\begin{equation}\label{dynamics}
        \dot{x_i} = u_i,
\end{equation}
where the agent velocity $u_i \in R^2, i\in \{1, 2, \dots, n\}$ is the control input, which is computed as follows
\begin{equation}\label{controlV}   
        u_i = - k_V(x_i - C(V_i)), \quad k_V > 0.
\end{equation}
It is proved in \cite{Jorge2004} that a local optimum of coverage objective $H$ will be achieved under \cref{controlV}, which makes the multi-agent system converge to the centroidal Voronoi configuration.

\section{Problem Formulation and Method Overview}\label{problemformulation}
\begin{figure}[ht]
     \centering
         \includegraphics[width=0.45\textwidth]{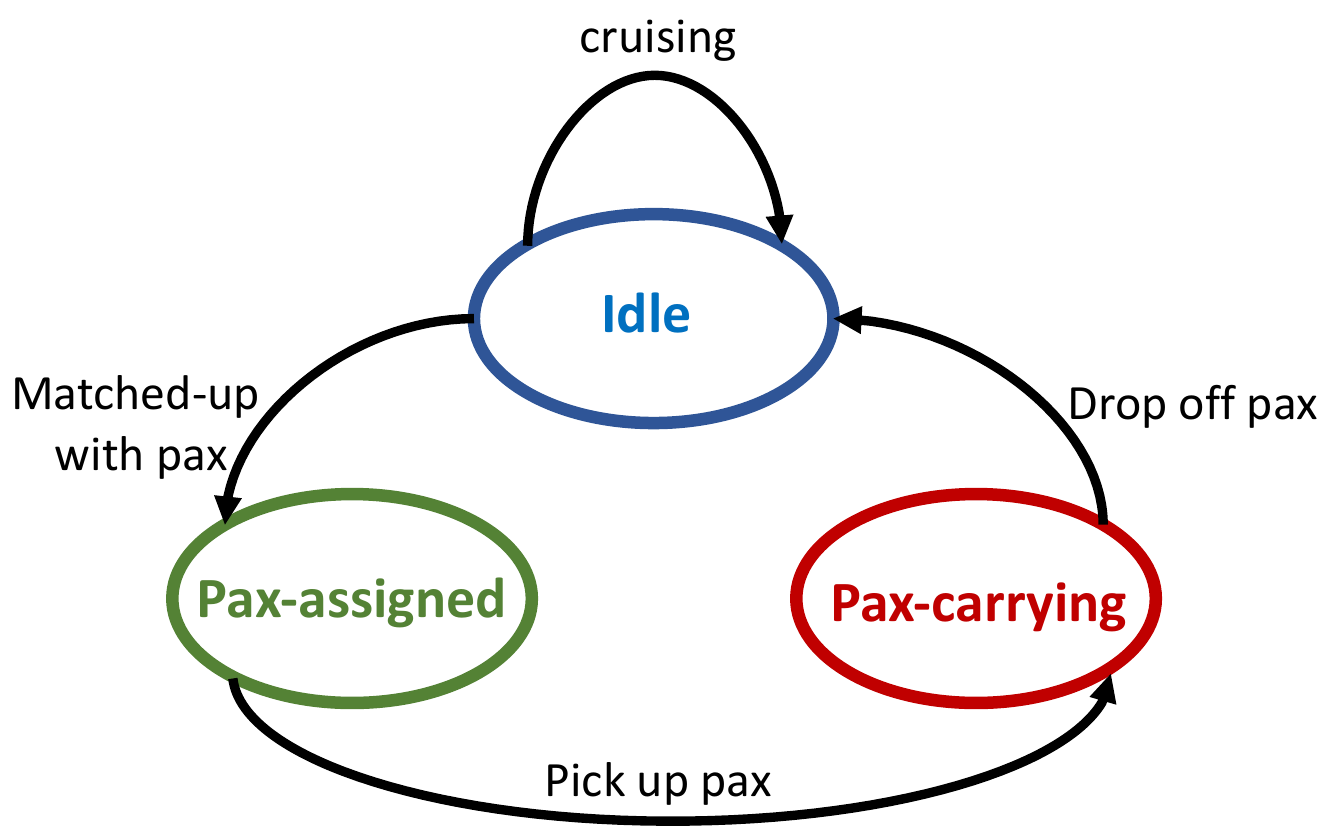}
         \caption{Finite-state machine schematic of an autonomous taxi. (`Passenger' is abbreviated as `Pax' for brevity.)}
         \label{fig:automaton}
       
     \end{figure}

We consider an AMoD system consisting of a fleet of taxis, all of which are autonomous vehicles (AVs) having identical capabilities of sensing, communicating, computing and controlling their own motion. Moreover, we assume that the AVs have an unlimited communication range and thus they can detect any other adjacent taxis accurately. This is feasible technologically since the TNC platform can collect their real-time GPS coordinates and occupancy status. In this work, we assume that each AV can serve only one passenger at a time (or a unit of passengers with the same origin and destination, thus being functionally the same as a single passenger). Therefore the AV has the following three occupancy states:
\begin{itemize}
    \item Idle (empty) vehicle: The vehicle is not carrying a passenger and is thus cruising/waiting for a passenger;
    \item Passenger-assigned: The vehicle is matched with a passenger and is moving to the passenger's origin position to pick up the passenger;
    \item Passenger-carrying: The vehicle has picked up its passenger and is moving to the passenger's destination position to drop off the passenger.
\end{itemize}
The transitions between these occupancy states are illustrated in \cref{fig:automaton} .

The focus of this work is on developing a control algorithm for the efficient spatial rebalancing of the set of idle vehicles of an AMoD fleet. The algorithm is based on the coverage control algorithm proposed for robotic systems in \cite{Jorge2004}, with the following intuitive reasoning: The pick-up task at a point $q \in V_i$ in the Voronoi cell $V_i$ should be executed by the vehicle closest to $q$, which is exactly the vehicle $i$ by definition of Voronoi cell. Furthermore, the density function $\phi$ of coverage control is utilized here to describe the spatial intensity of mobility demand in the AMoD operation scenario.

\subsection{Coverage control for Vehicle Rebalancing (CVR)}
By implementing \cref{centroidV}, the city area $\Omega$ is rasterized/discretized as a set of pixels, denoted as $\hat\Omega$. Let $|\hat A|$ stand for the number of pixel centers within the discretized area $\hat A$. Then the weighted centroid can be computed over the discretized Voronoi cell $\hat V_i$ (see Section 2.2 in \cite{Rougier2018}) as
\begin{equation}\label{centroidhatV}
 C(\hat V_i) = \frac{{\sum_{q \in \hat V_i}} \hat q\phi(\hat q)}{{\sum_{q \in \hat V_i}} \phi(\hat q)}, 
\end{equation}
where $\hat q$ are the pixel centers. Computing \cref{centroidhatV} has a complexity of $\mathcal{O}(|\hat V_i|)$, while the worst case can be  $\mathcal{O}(|\hat \Omega|)$.
% By implementing \cref{centroidV}, the covered area of each agent $V_i$ is rasterized and the weighted centroid  $C(V_i)$ is computed over each cell (see section 2.2 in \cite{Rougier2018}). Computing the centroid on continuous space has a complexity of $\mathcal{O}(|V_i|)$, where $|V_i|$ stands for the number of the centers of the rasterized grid in $V_i$, while the worst case can be $\mathcal{O}(|\Omega|)$. 
To limit the computational time, in contrast to \cite{Jorge2004}, we additionally assume that each agent can cover a circular area with a limited radius $r$ (shown as the gray circles in \cref{fig:final}). For instance, the area covered by vehicle $i$ (shown as the yellow disk in \cref{fig:final}) is as follows
\begin{equation}\label{circular}
    S_i(x_i, r) = \{q \in \Omega: \Vert q - x_i \Vert \leq r\}.
\end{equation}

At the same time, with the Voronoi partition, the actual area covered by vehicle $i$ is $W_i = V_i \cap S_i$. For instance, for the example situation in \cref{fig:final}, for vehicle $i$ (shown at the point $x_i$), $W_i$ is the same as $S_i$ (yellow area), while for vehicle $j$, $W_j$ is different to $S_j$ (as shown by the purple area). Computing the centroid of $\hat W_i$ then has a complexity of $\mathcal{O}( |\hat W_i|)$ (which is $\mathcal{O}(|\hat S_i|)$ in the worst case).

Similarly to the objective function (\ref{coverageobjV}) considered in \cite{Jorge2004}, we formulate the coverage objective function $H$ regarding $r$-limited Voronoi partition as follows,
\begin{equation}\label{coverageobjW}
    H(X,W) = \sum_{i= 1}^n \int_{q \in W_i }\Vert x_i - q \Vert^2\phi(q)dq,
\end{equation}
where $W = \{W_1, W_2, ... , W_n\}$.

For a region $W_i \subset \Re^2$, if we consider the probability density function $\phi$ as a mass density function, then the mass $M(W_i)$ is equal to ${\int_{W_i}} \phi(q)dq$.
% first moment $L_{W_i}$, 
The centroid $C(W_i)$ is defined by (\ref{centroidV}) and the polar moment of inertia $J(W_i, x_i)$ are given by:
% \begin{equation}\label{centroidW}
%  C(W_i) = \frac{{\int_{q \in W_i}} q\phi(q)dq}{{\int_{q \in W_i}} \phi(q)dq}, 
% \end{equation}
\begin{equation}\label{polar}
J(W_i, x_i) = {\int_{q \in W_i}}\Vert x_i - q\Vert^2\phi(q)dq.    
\end{equation}

By still assuming the dynamics in (\ref{dynamics}), the control law for each vehicle $i$ is defined as
\begin{equation}\label{controlW}
    u_i = -k_W(x_i - C(W_i)), \quad k_W > 0.
\end{equation}   
This control law guides every vehicle to move toward the centroid of $W_i$ with a speed $\frac{k_W}{||x_i - C(W_i)||_2}$. In our discrete-time implementation, the coverage objective (\ref{coverageobjW}) is renewed at each time step according to the current empty vehicle coordinates, and the control action (\ref{controlW}) is applied over a sampling period $\Delta T$. As per Appendix A, any positive value of $k_W$ guarantees a decrease of the objective function. Thus, even the value of speed may vary due to traffic congestion levels at different time steps, it does not violate the objective function's decreasing.
% This control law decreases the objective function (the proof can be found in Appendix \ref{appendix_control}), and each vehicle will move towards the centroid of $W_i$ with a speed $\frac{k_W}{||x_i - C(W_i)||_2}$. 
% % and tend to converge to the coverage-optimal configuration ( the proof can be found in Appendix \ref{appendix_control}). 
% In our discrete-time implementation, the coverage objective (\ref{coverageobjW}) is renewed at each time step according to the current empty vehicle coordinates, and the control action (\ref{controlW}) is applied over a sampling period $\Delta T$. 
For the sake of brevity, the specification of the time step $k$ is omitted in this section.

From the AMoD operation point of view, the control law operates the fleet by matching idle vehicle availability and trip demands via dispatching more empty vehicles to high-demand areas. Since it can be used for spatiotemporal deployment of idle vehicles of an AMoD fleet, we call the method Coverage control-based idle Vehicle Rebalancing (CVR) approach.

\textit{Remark 1:} For each vehicle, the control algorithm has two steps: 1) Using local information (positions of its spatial neighbors and itself), compute the covered area $W_i$ and weighted centroid $C(\hat W_i)$ (as an approximation of $C(W_i)$ according to \cref{centroidhatV}). 2) Move towards $C(\hat W_i)$. Therefore, the CVR method can be implemented in a distributed way (apart from the need for a central planner communicating demand density information to each vehicle). Moreover, the computation of the centroid scales linearly with $|\hat W_i|$ and therefore is suitable for real-time operations, enabling the rebalancing action for each agent to be updated in relatively short time intervals.

\begin{figure}[htb]
    \centering
\includegraphics[width=0.5\textwidth]{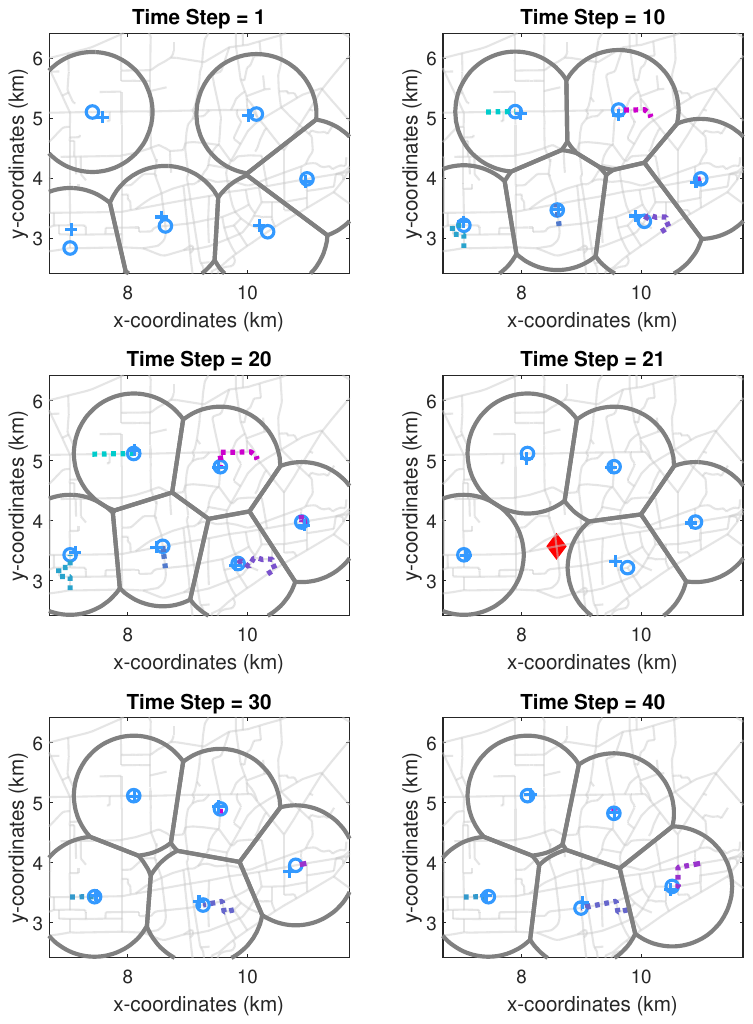}
    \caption{ A sequence of screenshots of vehicle rebalancing with CVR in action. `o' marks the current positions of the vehicles, while `+' indicates their rebalancing goal positions (i.e., Voronoi centroids). The trajectories of these vehicles are shown by dotted lines.}
\label{fig:fleet-changing}
\end{figure}

\textit{Remark 2:} \cref{controlW} controls the vehicle position to follow gradient descent flow (see Appendix \ref{appendix_control} for further details), which is not guaranteed to find the global optimum \cite{Jorge2004}.
The configuration is thus also affected by the initial position of agents. 
% Our goal is to improve service quality by using the proxy of coverage control. 
In contrast to previous works such as \cite{Jorge2004}, due to the number of idle vehicles changing over time (as they can be assigned to, or drop off a passenger anytime), here the coverage control problem is not static. Instead, the proposed method involves solving at each time instant a different coverage control problem, the dimensions/structure (due to changing idle fleet size) and data (due to changing demand intensity) of which can change.
% Thus, because of the time-varying nature of the setup, the coverage objective function value can be greater although the control algorithm is running. 
% Thus, because of the time-varying nature of the setup, the value of the coverage objective may increase although the control algorithm is running.
For instance, we present a scenario where the system starts with 6 vehicles in \cref{fig:fleet-changing},  At Time step = 21, one vehicle (represented by a red diamond) gets assigned to a passenger, and is no longer considered in the Voronoi partition. Consequently, the remaining vehicles adjust and aim to converge to a new configuration based on a fleet size of 5. The final figure illustrates that the vehicle configuration dynamically adapts according to the current availability of vehicles. Notably, the intended goal of the proposed method is not to monotonically decrease the coverage objective and eventually converge to a locally optimal configuration. Instead, the proposed method attempts to continually steer the idle fleet towards configurations with a lower coverage value, thus enabling the AMoD fleet to operate with greater potential to have short response times, thereby increasing service quality.

\subsection{Method Implementation}\label{CVR_AMoD}
 We use an undirected graph $G = (Q, E, w)$ to represent the city map, where $E$ is the set of road links and $Q$ is the set of vertices/nodes, i.e. intersections on city road, $N$ is the number of nodes in $G$, and the edge weights $w$ are the road length between two nodes. The proposed method is tested on an AMoD simulator replicating the urban road network of Luohu and Futian districts in Shenzhen, China. The network consists of $N = 1858$ intersections and $2013$ road links.

 \begin{figure}[!ht]
  \centering
  \includegraphics[width=0.46\textwidth]{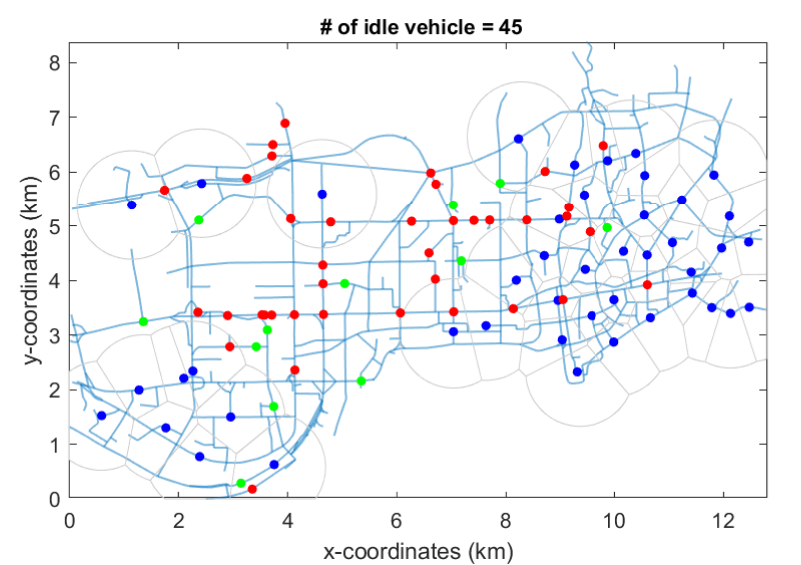}
  \caption{A snapshot of the simulator. The idle/empty, passenger-assigned, and passenger-carrying AVs are dots in blue, green, and red, respectively. A demo video is available on YouTube: \url{https://youtu.be/JlBs0CfuJ_c}.}\label{fig:sim_city}
\end{figure}

 \cref{fig:sim_city} shows a snapshot of the simulator. The blue dots stand for idle vehicles, and the gray contours denote the area covered by each vehicle (from the perspective of the coverage control objective function). In a practical setting, all vehicles should move along real urban roads and can only take turns or change their directions at the upcoming intersection. However, the calculated centroid $C(\hat W_i)$ may not be located on the roads that the vehicles can access. To circumvent this problem, instead of the computed exact centroid, we use the intersection/node closest to the centroid in the Euclidean metric as a simple approximation. Then, the implementation of the proposed CVR algorithm is provided in Algorithm \ref{AlgorithmMoD}.
\begin{algorithm}[htb] 
\SetAlgoLined

At a time step $T(k)$,\\
 \For{each idle vehicle $i$, $i \in \{1,2,\dots,n_{idle}(k)\}$ (with current position $x_i(k)$, upcoming intersection is $Q^{next}(i, k) \in Q$, and its current rebalancing destination $Q^{dest}(i, k)\in Q$.}    
        {   \textbf{Communicate:} Transmit its position and obtain the position information of its adjacent neighbors.\\
        	\textbf{Compute:}  Compute its Voronoi partition and the centroid of its $r$-limited Voronoi cell $C(\hat W_i, k)$. Find the nearest intersection $Q^{dest}(i, k + 1) \in Q$. \\
        	\textbf{Update:} $Q^{dest}(i, k + 1)\leftarrow C(\hat W_i, k)$

\textbf{Move:} The vehicle moves towards $Q^{dest}(i, k + 1)$ with speed given by \cref{MFD}.\\
        }
%  }
 \caption{Implementation of CVR.}\label{AlgorithmMoD}
\end{algorithm}

\section{Experiment Setup and Simulation Results}\label{Experiments}
Supply and demand imbalances commonly exist in MoD systems, which leads to inefficient operation. These imbalances can be caused by the asymmetry between trip origin and destination distributions and heterogeneous demand levels in different regions. Thus, we will discuss the Origin-Destination (O-D) imbalance and the imbalance magnitude in \cref{ODimbalance} and \cref{ImbalancePara}, respectively, and describe the passenger requests setting we use to test \cref{AlgorithmMoD} in \cref{simulationSetup}.
Then, the sensitivity analyses over different imbalance scenarios and fleet sizes are presented.
\subsection{O-D imbalance}\label{ODimbalance}    
\begin{figure}[h]
     \centering
     \begin{subfigure}[b]{0.45\textwidth}
         \centering
         \includegraphics[width=0.8\textwidth]{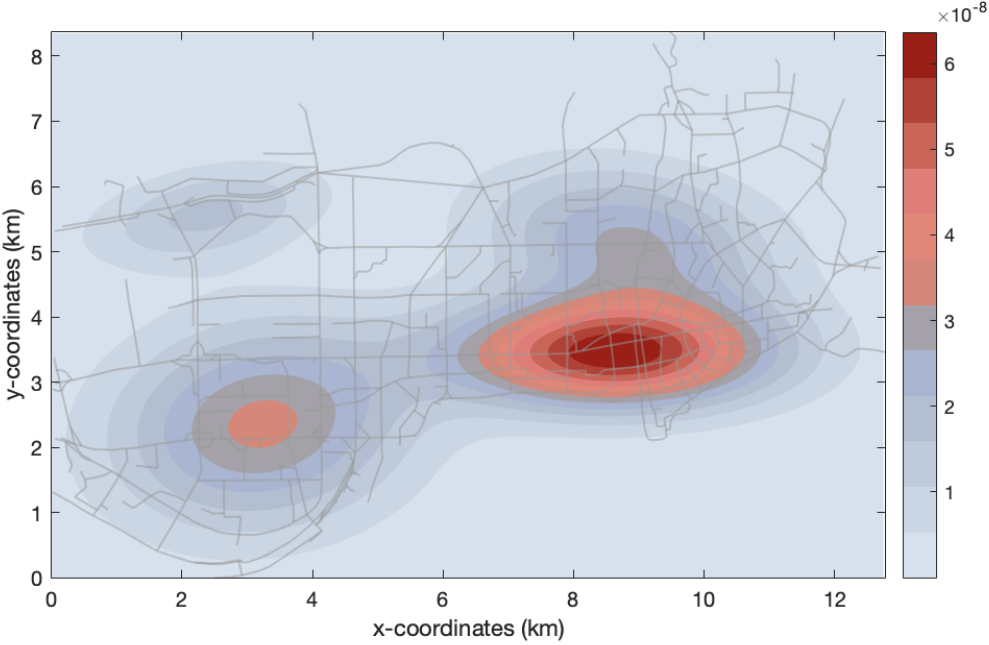}
         \caption{Origin distribution.}
         \label{fig:GMM_orig}
     \end{subfigure}
     \quad
     \begin{subfigure}[b]{0.45\textwidth}
         \centering
         \includegraphics[width=0.8\textwidth]{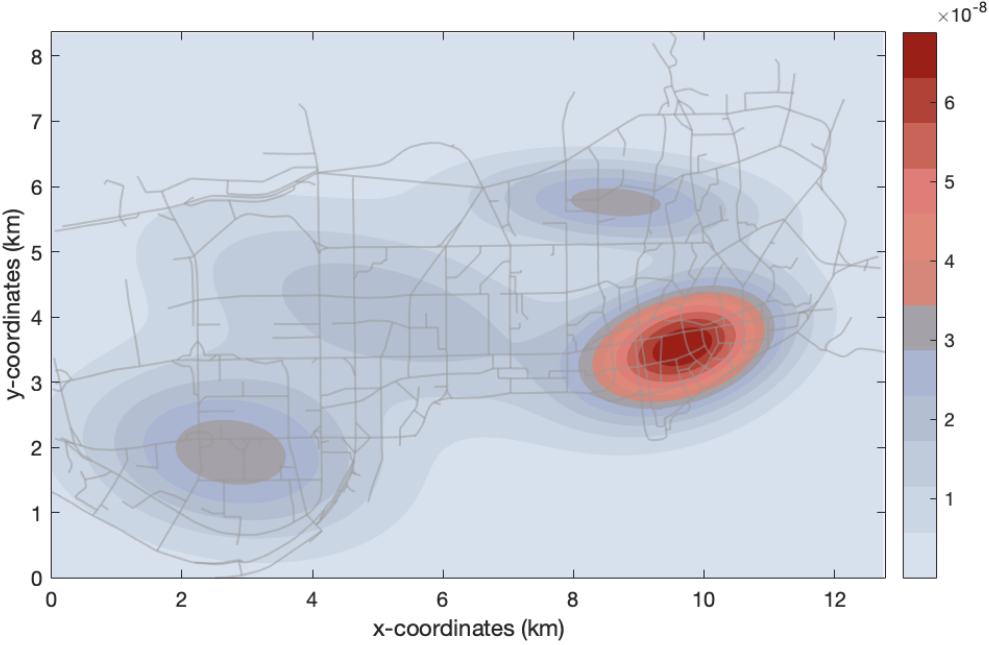}
         \caption{Destination distribution.}
         \label{fig:GMM_dest}
     \end{subfigure}
        \caption{Contour map of Estimated Gaussian mixture models for the trip origin and destination distributions. More orders begin/end at the darker colored/red areas.}
        \label{fig:GMM}
\end{figure}

The historical data contains 199,819 taxi trips (collected during a 24-hour period) from the city of Shenzhen, China, consisting of their origin-destination pairs. In order to present the impact of the spatial discrepancy between trip origin and destination distributions, we consider both distributions to be fixed in the following discussions.
Gaussian Mixture Models (GMMs) are used to estimate the spatial distributions of origins and destinations separately, as shown in \cref{fig:GMM}. The origin distribution represents where the passengers would like to start their trips. At the same time, this continuous GMM of origin distribution is used as a demand density function $\phi$ of the coverage control algorithm, which satisfies the condition $\int_{q \in \Omega} \phi(q)dq = 1$.
% \begin{figure}[h]
%      \centering
%      \begin{subfigure}[b]{0.35\textwidth}
%          \includegraphics[width=\textwidth]{graphics/contour_GMM_O.png}
%          \caption{Origin distribution.}
%          \label{fig:GMM_orig}
%      \end{subfigure}
%      \quad
%      \begin{subfigure}[b]{0.35\textwidth}
%          \includegraphics[width=\textwidth]{graphics/contour_GMM_D.png}
%          \caption{Destination distribution.}
%          \label{fig:GMM_dest}
%      \end{subfigure}
%         \caption{Contour map of Estimated Gaussian mixture models for trip origin and destination distributions. More orders begin/end at the yellow areas.}
%         \label{fig:GMM}
% \end{figure}

 Comparing \cref{fig:GMM_orig} and \cref{fig:GMM_dest}, it can be seen that these two distributions are different. Customer orders will gradually change the spatial distribution of supply (i.e., idle vehicle availability), since vehicles will move to respond to trip requests and tend to stay in the vicinity of drop-off (i.e., destination) points. Therefore, without any intervention, the supply distribution will shift towards the destination distribution (while ideally, it should match the origin distribution), resulting in severe spatiotemporal imbalances between supply and demand.

\subsection{Imbalance parameter}\label{ImbalancePara}
In the considered AMoD system, all requests start and end at intersections, i.e., nodes on the graph. Using the historical data, we define the total number of pick-up requests at node $i$ as $R^o(i)$, where $i = 1,2, ..., N$. Then the probability that a pick-up happens at node $i$ is as follows
\begin{equation}
    p^o(i) = \frac{R^o(i)}{\sum_{i=1}^{N}R^o(i)}.
\end{equation}
where $\sum_{i = 1}^{N} p_o(i) = 1$. Similarly, we can get the probability of drop-off at node $i$ as $p^d(i)$.

% In addition, to generate different demand scenarios under different levels of imbalance, here we introduce an imbalance parameter $\gamma$.
By defining the maximum of $p^o$ as $p^o_{max}$, we can write
\begin{equation}
    \Tilde{p}^o(i)= p^o_{max} - p^o(i).
\end{equation}
Then we normalize $\tilde{p}^o(i)$ over all nodes to have $\sum_{i = 1}^{N} \Tilde{p}^o(i) = 1$. We use $\Tilde{p}^o$ as a complement of $p^o$, in order to generate synthetic destination distributions from origin distributions through a parameter $\gamma \in [0,1]$ (representing O-D distribution balance) as follows
\begin{equation}
    p^d_{\gamma}(i) = \gamma \cdot p^d(i) + (1 - \gamma) \cdot \tilde{p}^o(i),
\end{equation}
where $\sum_{i = 1}^{N} p^d_{\gamma}(i) = 1$. When $\gamma = 1$ the generated destination distribution $p^d_{\gamma}$ is the same as the original one, i.e., $p^d$, while the smaller $\gamma$ is, the more imbalance there is between the generated destination distribution and the origin distribution. When $\gamma = 0$, $p^d_{\gamma}$ is the same as $\Tilde{p}^o$, which has a shape complementary to $p^o$ in that it generates the maximal O-D imbalance.

In addition, to test the effect of $\gamma$, we compute the Hellinger distance \cite{Hellinger1909} between $p^d_{\gamma}$ and $p^o$, which is a metric that takes values in $[0, 1]$ and it measures the degree of similarity between two probability distributions; when the distance is 0 the two distributions are identical and when it is 1 they are the furthest apart. The Hellinger distances between the generated destination distributions with different values of $\gamma$ and the origin distribution are shown in \cref{tab:hellinger}.

\begin{table}[h]
    \centering
	\caption{The Hellinger distances between $p^d_{\gamma}$ and $p^o$ for various values of $\gamma$}\label{tab:hellinger}
	\begin{tabular}{|c||c|c|c|c|c|}
		\hline
			 $\gamma$  &$ 0 $        & $ 0.25 $    & $ 0.5 $  & $ 0.75 $ & $ 1 $\\
			 
			 \hline
			 
			Hellinger distance &$0.2840$     & $0.2391$    & $0.1929$ &$0.1421$  & $0.0773$\\
			    \hline
	\end{tabular}
\end{table}

In the following sections, the origins of passenger demands are sampled from the origin distribution $p^o$ from historical data, while the destination distribution follows $p^d_{\gamma}$ generated by the above procedure.
\subsection{Simulator Setup and Performance Metrics}\label{simulationSetup}

% \subsection{Coverage Control algorithm for vehicle rebalancing}
% The arrival of passenger request is a stochastic process following certain distributions which is described in \cref{ODimbalance}. 
% The distribution of accumulation is considered to be homogeneous in the urban area of interest, thus a macroscopic fundamental diagram (MFD, see \cite{geroliminis2008existence}) is used to describe the relationship between the accumulation of vehicles $m$ (i.e., number of all vehicles consisting of private ones and AVs) and the space-mean speed $v$ (in m/s), as follows 
In our simulations, we assume that the congestion is homogeneous across the urban area of interest, thus a macroscopic fundamental diagram (MFD, see \cite{geroliminis2008existence}) is used to describe the relationship between the accumulation of vehicles $m$ (i.e., number of all vehicles consisting of private ones and AVs) and the space-mean speed $v$ (in m/s), as follows
\begin{equation}  \label{MFD}
    v(m) = \left\{
  \begin{aligned}
       &36e^{(\dfrac{-29m}{72000})} , &   m \leq 4320,& \\
       &6.31-2.33(m-4320),& 4320<m\leq 7200,& \quad\\
       &0,& m> 7200.&
  \end{aligned}
  \right.
\end{equation}
% where the values of numerical parameters are obtained from real taxi trip data as detailed in \cite{Ji2014}. 

The MFD is derived from traffic data, which can be collected through various kinds of sensors including loop detectors \cite{ZOCKAIE2018}, GPS on vehicles \cite{Sirmatel2020,SIRMATEL2021}, drone-mounted cameras \cite{PAIPURI2021} and so on. Then the values of numerical parameters are estimated via curve fitting or system identification. Vehicle moving speeds are then emulated based on this MFD, given the vehicle accumulation.
Note here, except extreme situations where network speed is zero (which is not included in the following simulation results), the magnitude of $v$ is always greater than $0$, which naturally satisfies the speed requirement in (\ref{controlW}). 

Similar to real MoD systems (such as Uber and Car2Go), the simulator match-up scheme obeys the `first-come-first-served' policy. A request can be described by a tuple including the origin, the destination, and the time when the request is sent out. The matching process is described in Algorithm \ref{alg:matching}. The passenger waiting time consists of two parts: Time spent waiting to be matched with a vehicle, and the time spent waiting to be picked up. In reality, people have a limited amount of tolerance to these, which we denote as $t_{mtol}$ and $t_{ptol}$ respectively. When passenger $j$ sends out a request at time $t_0(j)$, the system will search for available idle vehicles, calculate the estimated pick-up time $\hat t_p(j)$ (according to current moving speed) for the spatially closest vehicle, and then match the passenger with this vehicle if the $\hat t_p(j)$ value is appropriate.

\begin{algorithm}[ht]
\caption{Matching process in AMoD system}\label{alg:matching}
\textbf{Initialization:} $N_{req} = 0, N_{order} = 0$\\
\For{each passenger $j$ (with the request sent out at time $t_0(j)$)}{
        $N_{req} \gets N_{req} + 1$\\
        \While{$T(k) < t_0(j) + t_{mtol}(j)$ min}
                    {\textbf{Compute:} Find the spatially closest idle vehicle, compute pick-up time $\hat{t}_p(j)$\\
                    \textbf{if} $\hat t_p(j) - t_0(j) \leq t_{ptol}(j)$\\
                        \hspace{0.5cm} Match the passenger with this vehicle,\\
                        \hspace{0.5cm} $N_{order} \gets N_{order} + 1$,\\
                        \hspace{0.5cm} \textbf{Break}\\
                    \textbf{else} \\
                    \hspace{0.5cm} $k \gets k + 1$
                    }
}
\textbf{Return:} $N_{req}, N_{order}$
\end{algorithm}

If $\hat t_p(j) - t_0(j) > t_{ptol}(j)$, then the passenger keeps waiting for being matched for $t_{mtol}(j)$ min. At $(t_0(j) + t_{mtol}(j))$ min, if there are still no vehicles available that can respond to this passenger, she/he will cancel this request and leave the system. In this case, we assume that the passenger will choose to drive her/his private vehicle to travel, which will increase the accumulation $m$ in the road network, leading to congestion.

If $\hat t_p(j) - t_0(j) \leq t_{ptol}(j)$, this vehicle will be assigned to the passenger (see the green dots in \cref{fig:sim_city}). Once the passenger and a vehicle are matched, we assume that this request cannot be canceled anymore and the `request' becomes an `order' for the vehicle. The vehicle then travels toward the origin position of the passenger to pick her/him up. After picking up the passenger, it will travel to the passenger's destination (see the red dots in \cref{fig:sim_city}). All traveling routes between any two nodes are pre-computed by the Floyd-Warshall algorithm \cite{Floyd1962} which yields the shortest path between any two nodes on a graph. The passenger-assigned and passenger-carrying vehicles have no contribution towards the coverage objective, while the proposed control method will continuously balance the idle vehicles to gravitate towards the high-demand areas, improving idle vehicle availability.
% Therefore, there can be a quick response for current and future requests. 
Once a passenger arrives at her/his destination, the vehicle becomes idle again and joins the group of vehicles operated by the coverage control algorithm.

Considering that each vehicle might move in the network at various speeds at different times, we denote the time when passenger $j$ is actually picked up as $t_p(j)$, and the total number of successfully completed orders as $N_{order}$. The average waiting-to-be-picked time can be computed as:
\begin{equation}
      \bar t_w = \cfrac{\sum_{j = 1}^{N_{order}}[t_p(j) - t_0(j)]}{N_{order}}.
\end{equation}
We define \textit{completion rate} as the percentage of requests that are successfully completed as $N_{order}/N_{req}\times 100\%$. The $\bar t_w$ value has biases due to considering only the answered requests (i.e., orders) while ignoring the waiting-for-matching time of cancelled requests, so we introduce the average system time by giving a time penalty for the canceled requests as follows
\begin{equation}
    % \bar t_{sys} = \cfrac{\sum_{j = 1}^{N_{order}}(t_p(j) - t_0(j)) + (N_{req} - N_{order}) \cdot \beta \cdot t_{ptol}}{N_{req}},
        \bar t_{sys} = \cfrac{\sum_{j = 1}^{N_{order}}(t_p(j) - t_0(j)) + \sum_{j=1}^{N_{req}-N_{order}} \beta \cdot t_{ptol}(j)}{N_{req}},
\end{equation}
where $(N_{req} - N_{order})$ is the number of canceled requests for the whole simulation, $\beta$ is a weight parameter representing the time value penalty for cancelled orders (with $\beta > 1$). Note that the value of $\beta$ does not affect the positive linear relationship between $\beta$ and $\bar t_{sys}$, and it is chosen as $\beta = 1.5$ for the simulations. We choose a constant tolerance time for all passengers that $t_{mtol}(j) = 1$ min and $t_{ptol}(j) = 5 $ min. Further simulation studies (omitted here for brevity) revealed that the system performance stops improving for $r$ values larger than $r = 1000m$, thus we choose this value for the simulations.

\subsection{Results and Analysis}
\begin{figure}[ht]
  \centering
  \begin{subfigure}[b]{0.48\textwidth}
        \centering
         \includegraphics[width=0.9\textwidth]{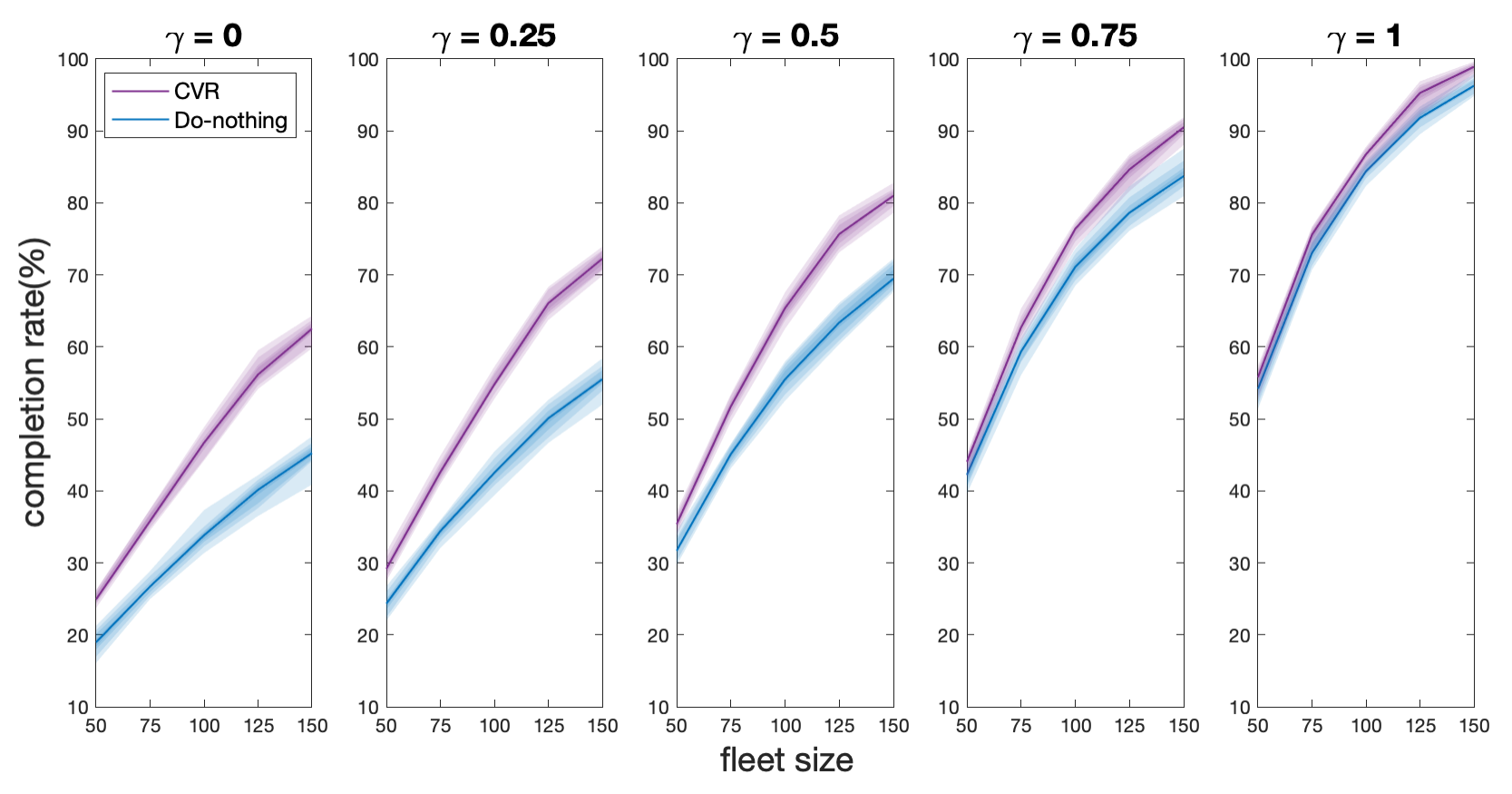}
         \caption{Request completion rates}
         \label{fig:comp}
     \end{subfigure}
     \begin{subfigure}[b]{0.48\textwidth}
        \centering
         \includegraphics[width=0.9\textwidth]{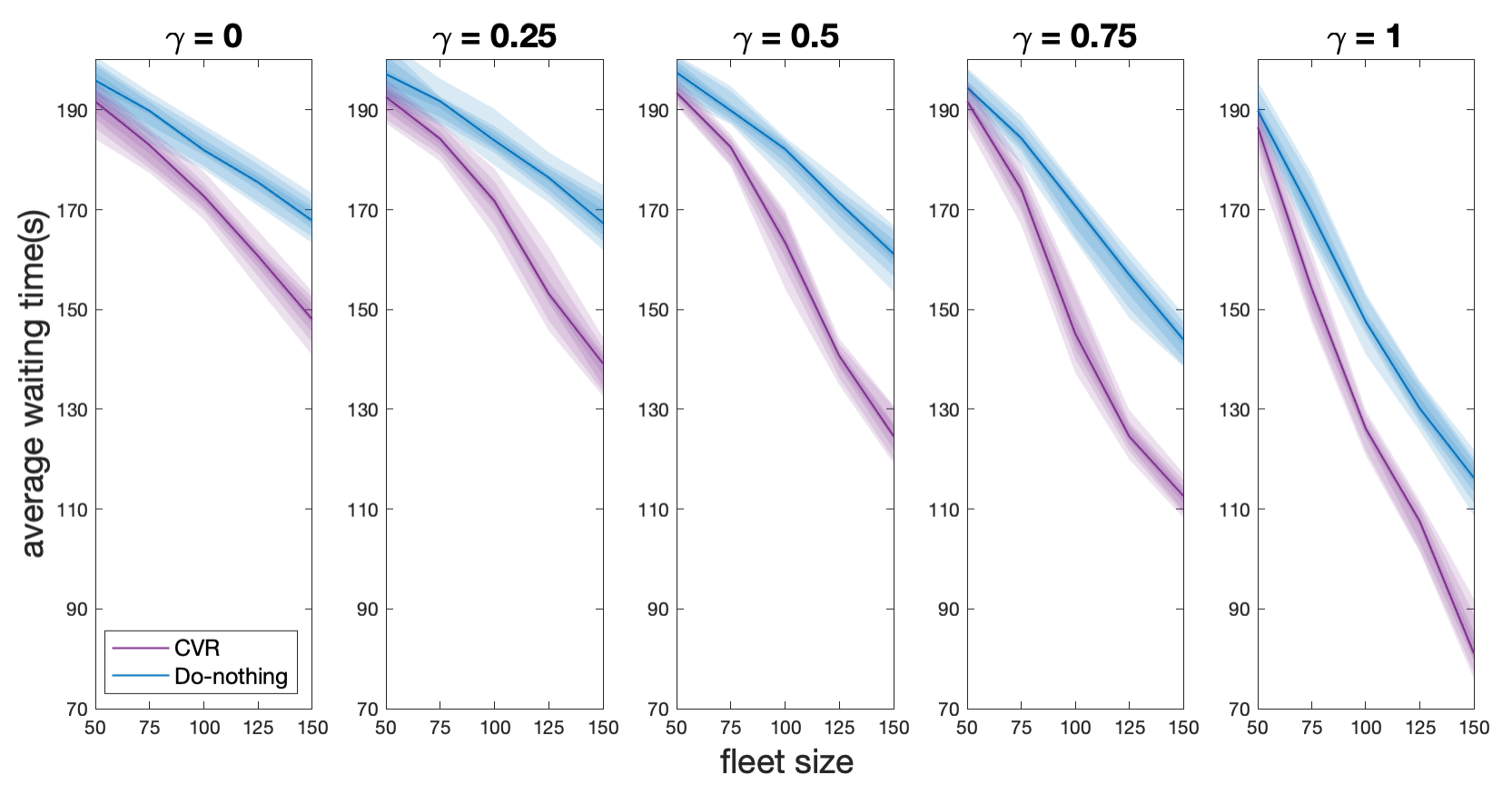}
         \caption{Average waiting times $\bar t_w$}
         \label{fig:wt}
     \end{subfigure}
       \begin{subfigure}[b]{0.48\textwidth}
       \centering
         \includegraphics[width=0.9\textwidth]{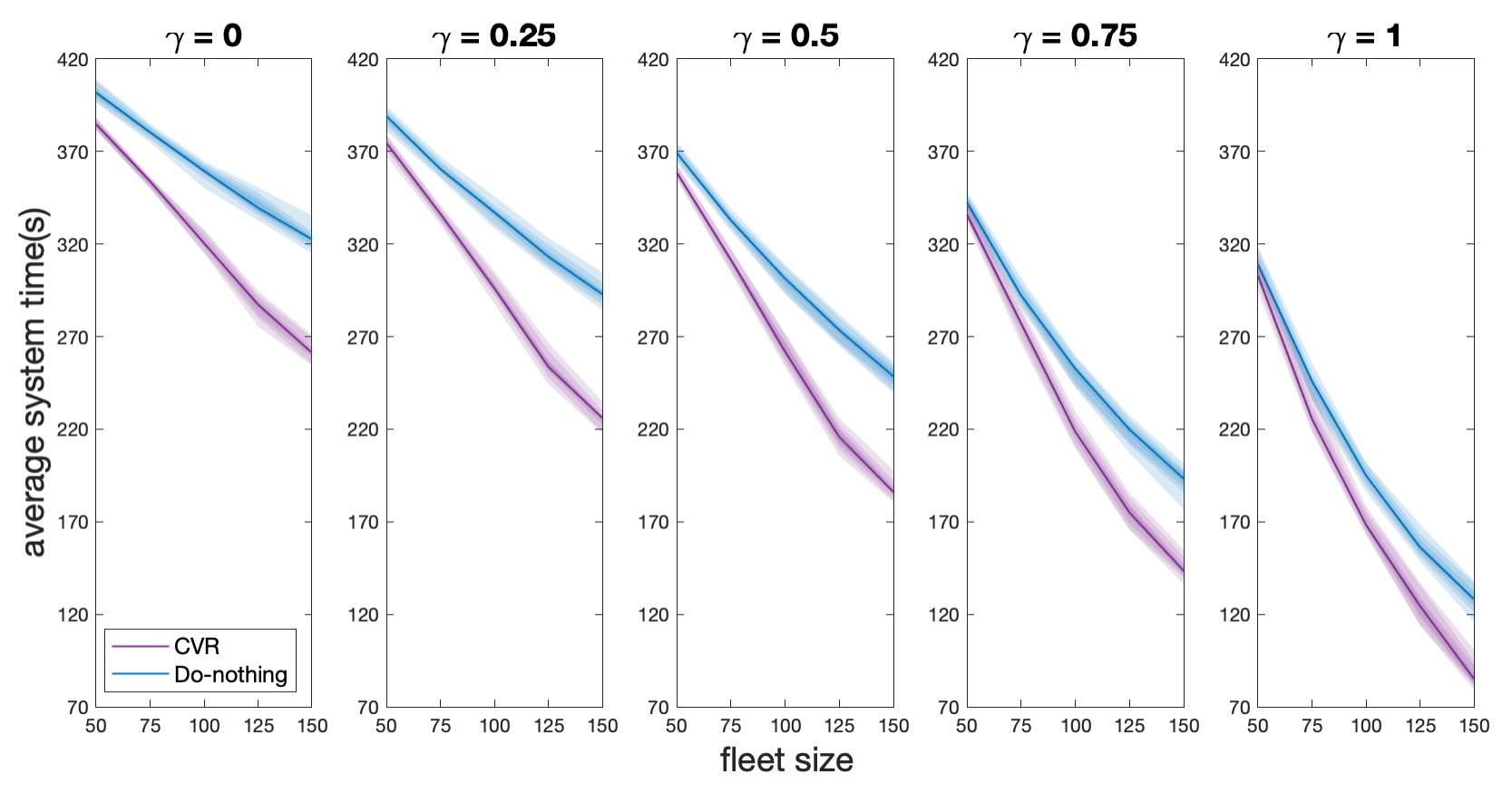}
         \caption{Average system times $\bar t_{sys}$}
         \label{fig:sys}
     \end{subfigure}
  \caption{Comparison for various $\gamma$ and fleet size values.}
  \label{fig:gamma_fleet}
  \vspace{-0.5cm}
\end{figure}

In order to illustrate the influence of trip origin and destination imbalances, we generate synthetic scenarios with different values of imbalance parameter $\gamma$. Besides the spatial distribution, the arrival process of the passenger requests follows a Poisson distribution with piece-wise constant rates with a low-high-low demand profile where each period lasts for 1 hour. During the first and third hours (i.e., the low-demand periods), around 600 requests are issued per hour, while the arrival rate doubles during the second hour (i.e., the high-demand period). In total, around 2400 demand requests are introduced in a 3-hour simulation, and we assume the spatial distributions of requests are fixed. 
% A spatial-temporal varying $phi$ can be collected from 
The baseline method, where vehicles stay at the destination of their last order until they are matched with their next passengers, is denoted as the \textit{`Do-nothing' policy}. The proposed method CVR is compared with the Do-nothing policy using a set of $30$ randomly generated simulation scenarios, and the results are provided in \cref{fig:gamma_fleet}, showing the request completion rates, average waiting times and system times, with varying values of $\gamma$ and fleet size, for the two methods. The solid lines present the mean value for the $30$ runs, while percentiles are shown in shades for varying degrees ($25\%, 50\%, 75\%$, and $90\%$, respectively). The coverage control algorithm is operated every $\Delta{T} = 10$ s throughout this paper, unless otherwise specified.

According to \cref{fig:gamma_fleet}, for all simulations,
% using various values of the origin destination demand imbalance parameter $\gamma$, 
our method can improve system performance by yielding lower waiting and system times, and is able to serve more trips. These results indicate the potential of the proposed method in improving AMoD system performance by allocating more vehicles around high-demand areas and dynamically rebalancing their positions after dropping off passengers. When $\gamma = 1$, both methods do well due to the origin and destination distributions being similar and the situation requires little rebalancing effort.
% , where the Do-nothing approach is a good enough choice. 
Even if there are no significant differences between the two methods regarding the completion rate (i.e., only an improvement of around $3\%$), our method shows substantially lower waiting and average system times, especially with large fleet sizes. With lower $\gamma$ values, the performances of both methods decline as expected. With an increased imbalance, there will be more orders with origins and destinations further away from each other, causing the coverage control algorithm to require more time to steer the vehicles from their last destinations towards the high-demand areas. However, overall the simulations show that the proposed method yields better results than the baseline.

From the results, considering the varying fleet size values, it can be seen that the performance metrics improve for both methods with increased fleet size as expected. Furthermore, the proposed method can achieve similar or better performance with smaller fleets. For example, in \cref{fig:wt}, for $\gamma = 0.5$ and a fleet size of $100$, the proposed method can yield an average waiting time $\bar t_w$ of around $160$ s, indicating an improvement of about $11\%$ compared to the baseline of $180$ s. In other words, to achieve a $\bar t_w$ around $160$ s, the baseline requires a fleet size of about 150, which is 50 vehicles more than what the proposed method needs to achieve the same performance. Given that larger fleets are problematic due to increased costs for the TNCs and can also create additional congestion in the urban network, the capability of the proposed method in achieving good performances with smaller fleets is a desirable feature
% a critical point indicating its strong potential 
for practical impact.

\section{Extension and Comparisons}\label{Extension}

Although CVR on continuous space is easy to implement, the continuous space assumption suffers from several disadvantages: Firstly, the coverage control algorithm requires a continuous demand density function. Estimating a continuous density function from discrete historical trip data can introduce errors. Furthermore, areas accessible by a vehicle are confined to road networks in practical scenarios. Thus, the coverage region of one vehicle is non-circular and non-convex in shape (which can be thought of as the intersections and roads surrounding the vehicle), in contrast to the circular disk assumed in the original coverage control method (as described in \cref{Preliminary}). Furthermore, the centroid calculated with the continuous Voronoi partition might be outside of any intersection on the city map, thus the closest intersection is used as an approximate position of the exact centroid. These steps inevitably introduce approximation inaccuracies. To overcome these problems, as the city map can be modeled as a graph, the original coverage control algorithm can be extended to coverage on graph algorithm (following \cite{Durham2012}) using the graph Voronoi partition method of \cite{Martin2000}.

\subsection{Problem Reformulation}\hfill

The urban road network is modeled as an undirected graph $G = (Q, E, w)$. 
% By only considering neighbor nodes that are within $r'$ limits, the computation time is acceptable.
For the vehicle $i$, whose current position is $x_i$, the Voronoi tessellation of graph is given by the cells \cite{Martin2000}
\begin{equation}
V_i^{G}(x_i) = \{q\in Q: d(x_i, q) \leq d(x_j, q)), \forall i \neq j \},
\end{equation}
where $d(A,B)$ stands for the shortest distance between nodes $A$ and $B$ in graph $G$.

Considering concerns related to computational complexity, similar to the continuous case where an $r$-limited Voronoi cell is used to depict the covered area of each agent, the graph coverage control scheme is implemented by bounding the graph Voronoi cell within $r^G$.
Taxi traveling on dense urban roads can be approximated by motion on a grid map, to relate the coverage on the graph to continuous space, here we choose $r^G = \sqrt{2}r$, which measures the coverage radius $r$ in \cref{circular} on the grid map by Manhattan Distance, i.e., the shortest path that a taxicab would take between city intersections), which is widely used in path planning \cite{Yang2020}.
% the sum of absolute difference of the components of two vectors
For the covered nodes of each idle vehicle $i$, we only consider the nodes within its graph Voronoi cell which are not further than $r^G$ limits from its position $x_i$. Therefore, we obtain the `covered area' of each idle vehicle on the graph as
\begin{align}
    S_i^{G}(x_i, r^G) = \{q \in Q: d(q, x_i) \leq r^G\}.
\end{align}
The $r^G$-limited graph Voronoi cell of vehicle $i$ is then $W_i^G = V_i^G \cap S_i^G$. A discrete version of the objective function \cref{coverageobjW} can be formulated as:
\begin{equation}\label{obj_G}
    H(X,W^{G}) = \sum_{i=1}^n J^G_i(x_i, W_i^G) = \sum_{i=1}^n \sum_{q\in W_i^G}d(x_i, q)^2\phi^G(q),
\end{equation}
where the demand density $\phi^G$ is a probability mass function over the nodes in $G$, which is the same $p^o$ introduced in \cref{ODimbalance}. It satisfies
\begin{equation}
    \sum_{q\in Q}\phi^G(q) = 1.
\end{equation}

According to \cite{Joseph2011}, the centroid of $r^G$-limited graph Voronoi can be computed by an integer optimization problem as
\begin{equation}\label{centroidG}
    C(W_i^G) = \arg\min_{q\in W_i^G} J_i^G(x_i, W_i^G).
\end{equation}
If the feasible solution of \cref{centroidG} is not unique, we randomly select one element from the solution set.

\begin{figure}[h]
     \centering
         \includegraphics[width=0.45\textwidth]{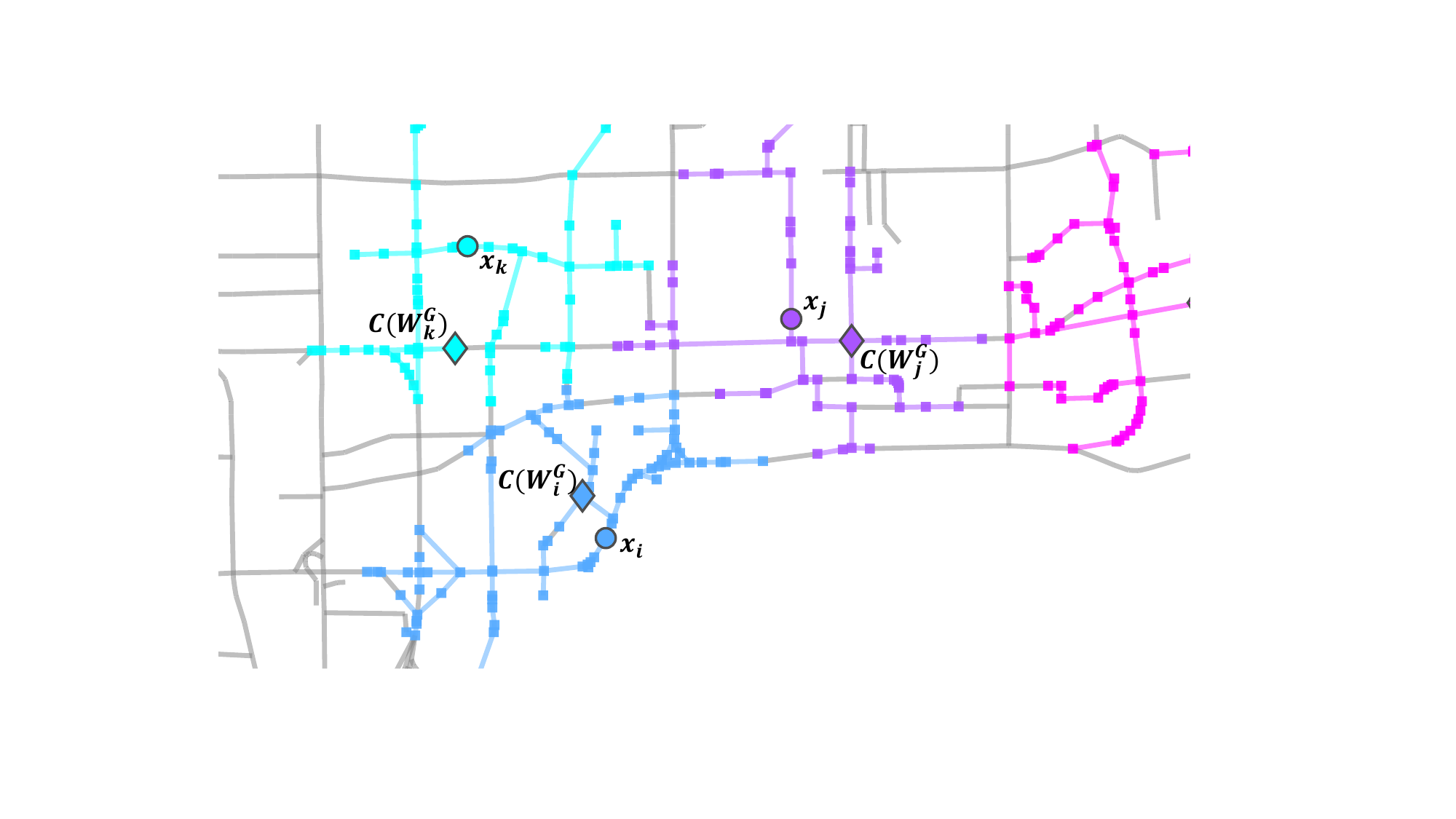}
         \caption{A snapshot of partition on graph of urban road. For each vehicle (round markers), the square markers in the same color around it stand for the graph nodes which are closer to it than to any other within a distance of $r^G$. The diamond marker illustrates the position of its centroid where this idle vehicle should move to.}
         \label{fig:graphVoro}
     \end{figure}

\cref{fig:graphVoro} is a snapshot of the $r^G$-limited graph Voronoi partition. For instance, for the vehicle in $x_i$, the covered nodes are shown as blue squares, and the graph centroid is denoted as $C(W_i^G)$. This idle vehicle $i$ will travel towards $C(W_i^G)$ following the shortest path calculated by the Floyd-Warshall Algorithm. We name this control algorithm as CVR-graph.
    %  \begin{figure}[h]
    %  \centering
    %      \includegraphics[width=0.3\textwidth]{graphics/graphPartition.pdf}
    %      \caption{Illustration of Voronoi partition on graph. Here we show a 3-vehicles case for visualization. For each vehicle (blue dots), the square markers around it stand for the nodes which are closer to it than to any other within a distance $r^G$. The diamond marker illustrates the position of its centroid where this idle vehicle should move towards.}
    %      \label{fig:graphVoro}
    %  \end{figure}

% \subsection{Results and Analysis}
% \begin{table}[h!]
%   \begin{center}
%     \caption{Performance matrics of CVR-graph}\label{table_graph}
%      \vspace{2ex}
%     \begin{tabular}{l|c|c|c|c} % <-- Alignments: 1st column left, 2nd middle and 3rd right, with vertical lines in between
%     \hline
%       \textbf{   }&\textbf{completion rate} & \textbf{waiting time} & \textbf{system time} & \textbf{empty travel trip}\\
%       \textbf{   }&{($\%$)} & {($s$)} &{($s$)}& {($km$)}\\

%       \hline
%       CVR continuous    & 82.6    &127.0  &183.2  & 4804.8    \\
%       \hline
%      CVR graph & 81.6 & 125.6 & 185.3 & 4894.3\\
%       \hline
%     \end{tabular}
%   \end{center}
% \end{table}

\textit{Remark 3:} 
% For a non-tree graph, solving \cref{obj_G} can be seen as a p-median problem in location optimization which is NP-hard\cite{Yun2012}. 
In contrast to the CVR on a continuous space, there is no need to estimate the demand density with CVR-graph. The partition on the graph can cover non-convex regions as required for practical coverage. The centroid can also be directly obtained by \cref{centroidG} without approximation. The disadvantage of the CVR-graph method is the increased computational effort. On the graph of urban roads, the edge weights are not the same as they reflect the length of road segments. For this non-tree graph, computing the centroid is the most time-consuming part, and the core of it is computing the distances between one node $q$ and all the other nodes in $W_i^G$ (one-to-all distances) which requires $\mathcal{O}(|W_i^G|log|W_i^G|)$ by Dijkstra's algorithm. Please note, by slight abuse of notation, in this section, $|A^G|$ stands for the number of all nodes on graph `covered' by $A^G$.

\subsection{Results and Comparative Analysis}
We introduce a comparison of our approaches with a fleet rebalancing strategy detailed in \cite{AlonsoMora2017A}. This strategy focuses on relocating idle vehicles to intersections where there are pending or unassigned requests. The strategy operates through linear programming (LP) that aims at minimizing the total travel time between pairs of vehicles and requests, subject to the condition that either all pending requests or all idle vehicles are allocated. While our proposed methods, CVR and CVR-graph, control the fleet in a distributed manner, the LP strategy from \cite{AlonsoMora2017A} operates in a centralized way and requires real-time access to information regarding all unassigned requests. 

An experiment is carried out under origin-destination trip demand imbalances (with $\gamma=0.5$) and a fleet size of 150. Around 2400 requests are introduced in a 3-hour simulation to compare the performances of CVR on the continuous map and on the graph. The control sampling time of all policies is set as $\Delta T = 10$ s. The performance metrics are listed in \cref{table_graph}.
% But it should be mentioned that, it is not that easy to develop a fleet sizing algorithm on a graph as we did for CVR because of the computation time.
\begin{table}[ht]
  \begin{center}
    \caption{Performance metrics of different policies}\label{table_graph}
     \vspace{2ex}
    \begin{tabular}{|c||c|c|c|} % <-- Alignments: 1st column left, 2nd middle and 3rd right, with vertical lines in between
    \hline
       \textbf{   }&\textbf{completion rate} & \textbf{waiting time} & \textbf{system time} \\
       \textbf{   }&{($\%$)} & {($s$)} &{($s$)}\\
      \hline
   
    CVR-graph & 82.7 & 125.7 & 181.9\\
      \hline
      CVR   & 82.7    &127.4  &183.1   \\
      \hline
LP\cite{AlonsoMora2017A}   &80.8  &150.8  &208.1\\

      \hline
      Do-nothing   &72.3    &157.6  &238.5   \\
      \hline
    \end{tabular}
  \end{center}
\end{table}

Compared with both LP and Do-nothing policies, CVR-graph yields better results: Answering more requests, at the same time, gives a clear reduction in average waiting and system time. The results demonstrate CVR-graph can help to improve the efficiency of the AMoD system. 

\begin{figure}[htpb]
    \centering
    \includegraphics[width=0.48\textwidth]{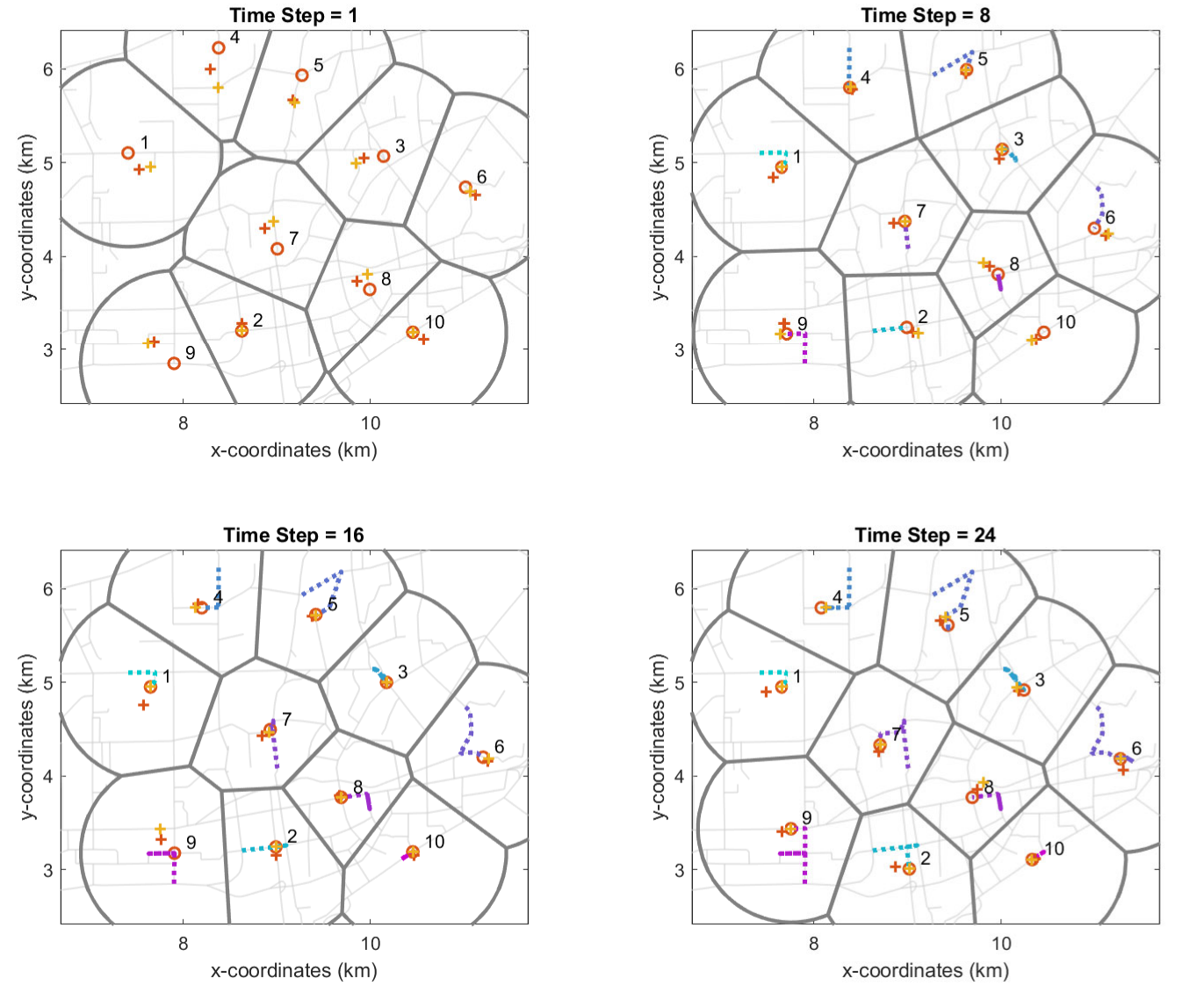}
    \caption{A sequence of screenshots with CVR in action. `o' are the current positions of vehicles, red `+' show their exact Voronoi centroids, and orange `+' are approximate centroids (true rebalancing destination/goal positions). The dotted lines represent their trajectories.}
    \label{fig:seq_CVR}
\end{figure}

\begin{figure}[htpb]
    \centering
    \includegraphics[width=0.48\textwidth]{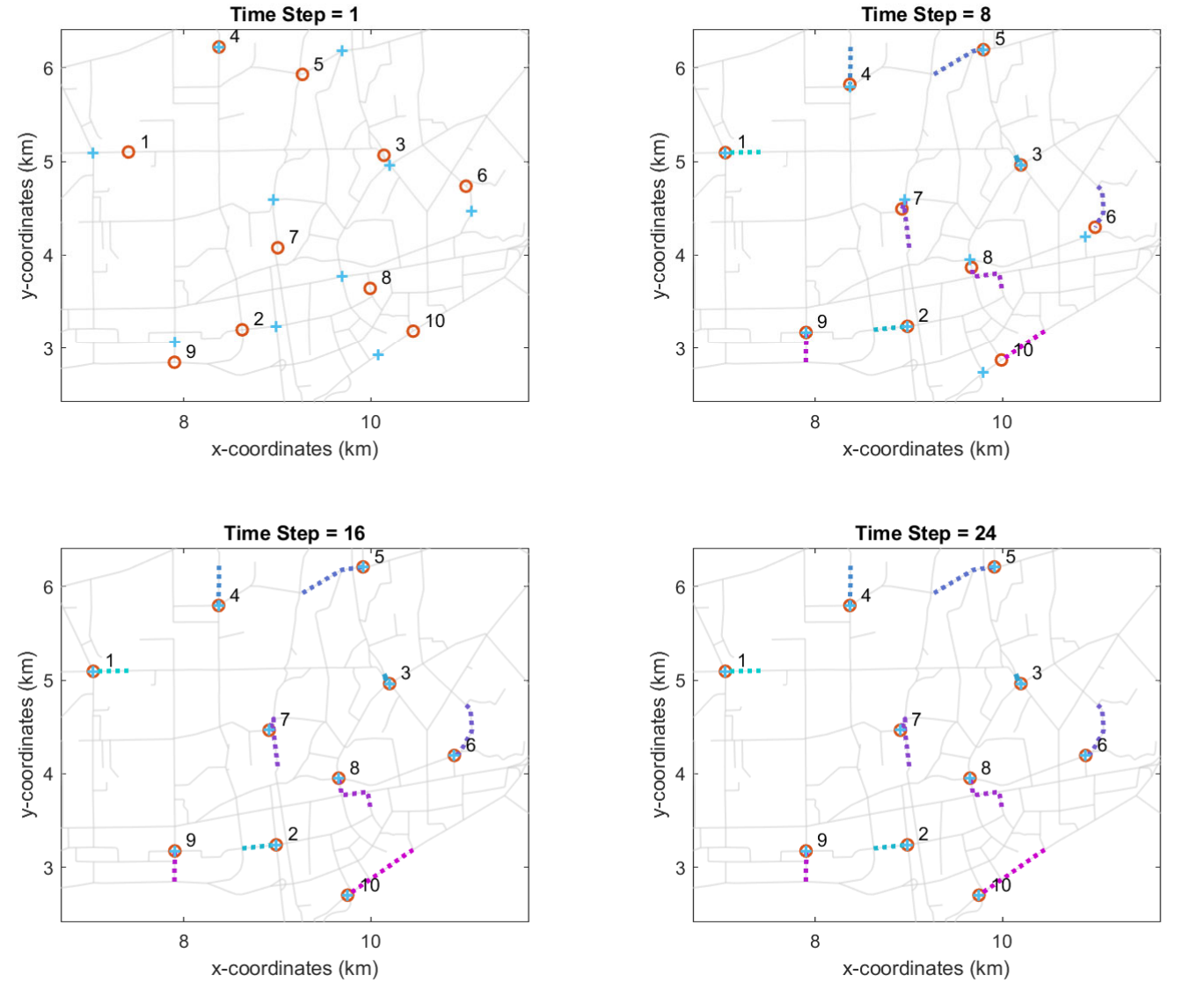}
    \caption{ A sequence of screenshots with CVR-graph in action. `o' are the current positions of vehicles, blue `+' show their centroids of graph Voronoi cell. The dotted lines represent their trajectories.}
    \label{fig:seq_CVR_graph}
\end{figure}

Furthermore, we compare CVR-graph to CVR proposed in \cref{problemformulation}.
In \cref{fig:seq_CVR}, the trajectories of vehicles under CVR are depicted. The vehicles move towards high-demand regions. Notably, Vehicles 9 and 10 show back-and-forth movements, potentially due to centroid approximation errors. For instance, while the actual centroids (red `+') for Vehicle 9 at time steps 8 and 16 are proximal, both of them lie outside intersections on the graph. Then the closest intersections (orange crosses) are used as approximate centroids, i.e., their rebalancing destinations. It results in a change of destination, causing Vehicle 9 to move westward first (as seen at Time Step 8), reverse its course (Time Step 16), and then head west again (Time Step 24). A similar pattern is observed with Vehicle 10. In \cref{fig:seq_CVR_graph}, the trajectories of vehicles operating under CVR-graph are presented. In contrast to CVR, there is no observed oscillatory movement, thanks to the exact centroid computation.

According to the performance metrics as listed in \cref{table_graph}, the CVR-graph is seen to yield similar performance to CVR, with a slightly shorter waiting time. This might be attributed to the studied city map, where the distribution of graph nodes is dense and roughly uniform. Consequently, approximating this road network with a continuous space turns out to be a reasonable approximation, as can be evidenced by \cref{fig:seq_CVR} showing that the exact centroids (red) and the approximate centroids (orange) are close. Although CVR and CVR-graph yield similar performances in studied dense road networks, discrepancies between these two methods could become more significant when vehicles are operated in dedicated lanes or utilize only a specific subset of the road network.

% \vspace{-0.3cm}
To sum up, CVR involves approximations that can lead to inaccuracies and potentially unnecessary oscillatory movements. In contrast, CVR-graph, though computationally relatively more intensive, provides more accurate and practical guidance for vehicle deployment.

\begin{figure}[h]
    \centering
    \includegraphics[width=0.5\textwidth]{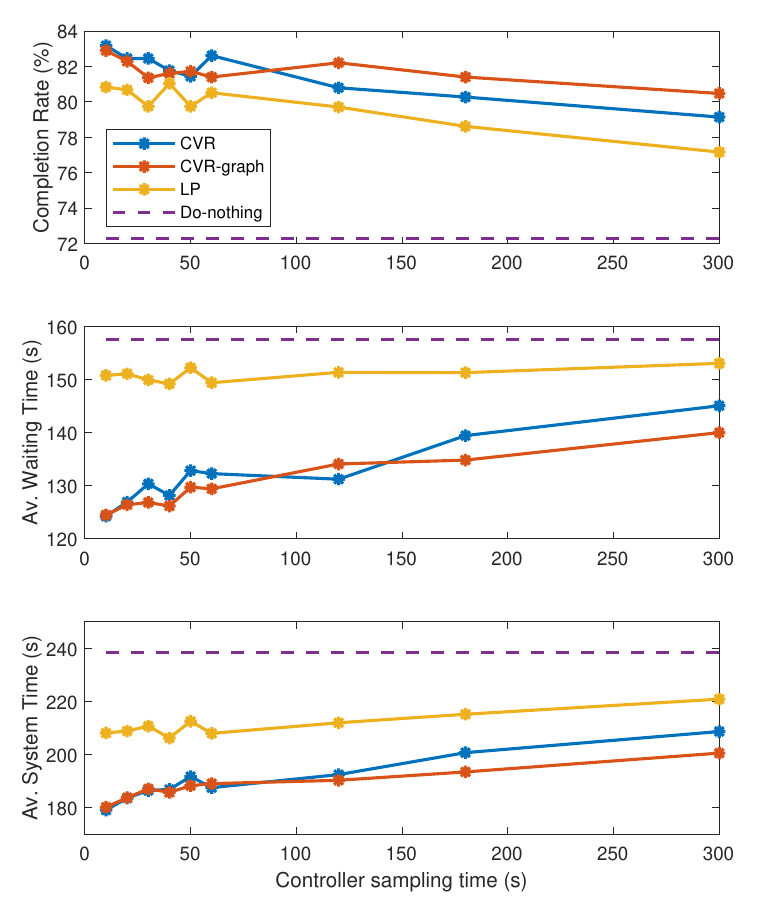}
    \caption{ Performances of CVR, CVR-graph, LP policy and Do-nothing policy as functions of controller sampling time.}
    \label{fig:CVR-LP}
\end{figure}
% \vspace{-0.5cm}

\subsection{Performance across different controller sampling time}

We test the performance of our methods, CVR and CVR-graph, and the LP policy for a set of control input update period (i.e., $\Delta T$, controller sampling time) values ranging between $10$ and $300$ seconds. The results are shown in \cref{fig:CVR-LP}. 

It can be seen that performance deteriorates for all methods as the controller sampling time increases, because the system becomes less responsive and fewer vehicles are rebalanced to high-demand regions as expected. Our methods consistently outperform the LP policy, addressing more requests and reducing waiting times across all tested scenarios. 
LP policy bases its control actions on the unfulfilled orders from the previous control sampling period, thus a longer sampling time results in less timely actions. For short controller sampling times ranging from 10s to 60s, CVR and CVR-graph show comparable completion rates, average waiting, and system time. However, when the sampling time is extended, CVR-graph demonstrates superior performance over CVR.
While the performance metrics of answer rate and average waiting time for CVR and CVR-graph deteriorate as the sampling time increases, they remain consistently better than those of the LP policy and Do-nothing policy.
\section{Active idle fleet size adaptation}\label{fleetsizing}
% \subsection{Motivation}

Outside of rush hours, there might be many idle vehicles circulating without serving any passengers. Furthermore, the proposed coverage control algorithm might force idle vehicles to execute many short back-and-forth movements due to small changes in idle fleet size inside short time intervals. The resulting empty kilometers traveled cause fuel waste, air pollution, and congestion. We therefore investigate two different methods for selecting a subset of idle vehicles not to be relocated. 
% We call these \textit{`not-relocated idle vehicles'},
By contrast, the rest of the idle vehicles (which are actively repositioning themselves according to the coverage control law) are termed \textit{`active idle' vehicles}. The goal of these methods is to reduce empty travel distances caused by mostly unnecessary small rebalancing actions.

There exist two questions in the idle fleet size selection problem:
1) How many vehicles should be relocated for different demand levels, and 2) which ones should be chosen for relocation?

Firstly, for each idle vehicle $i$ (with current position $x_i$), the polar moment of inertia of its covered area is given by \cref{polar}. Similarly, we can define the polar moment of inertia of its Voronoi cell as
\begin{equation}
J(V_i, x_i) = {\int_{q \in V_i}}\Vert x_i - q\Vert^2\phi(q)dq.   
\end{equation}

\begin{figure}[ht]
  \centering
  \includegraphics[width=0.48\textwidth]{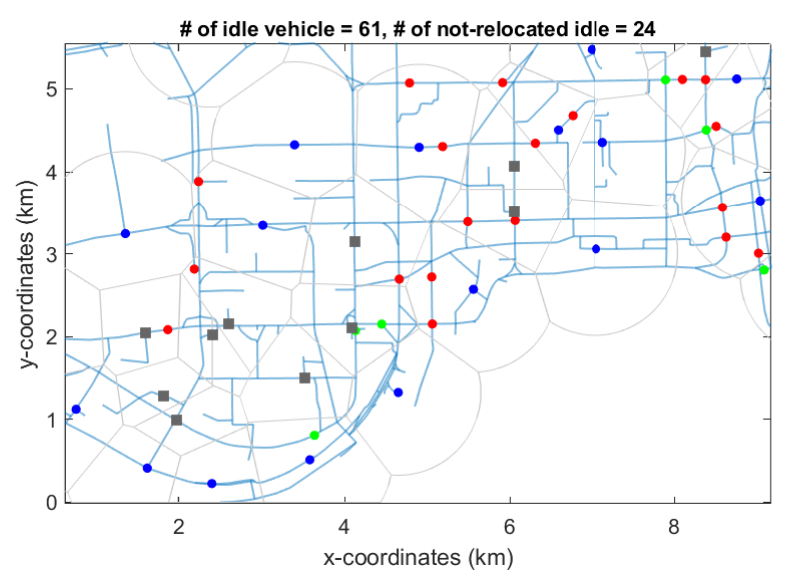}
  \caption{A partial screenshot of the simulator. The `active idle' vehicles are shown as blue dots, while the `not-relocated idle' vehicles are shown as gray squares.}\label{fig:sim_city_not}
\end{figure}

To guarantee service quality, we prefer to select vehicles that are located in high-demand areas and ask them to hold their positions. In a map with non-uniform density, with the proposed coverage control scheme, more idle vehicles will gather around high-demand regions (see \cref{fig:sim_city}). This implies that a vehicle located at a high-demand area usually has a higher value of $J(W_i, x_i)/J(V_i, x_i)$, since owing to the nature of Voronoi partition, $W_i$ will cover more area of the whole $V_i$. Note that both `not-relocated idle' and `active idle' vehicles should be considered for Voronoi partitioning (the contours of their covered areas are shown in gray in \cref{fig:sim_city_not}).  
\vspace{-0.5cm}

\subsection{Active idle fleet size adaptation controllers}
We introduce two different approaches to determine how many vehicles should not be relocated: 1) CVR-$\alpha$ simply chooses $\alpha$ percent of idle vehicles, 2) CVR-PI uses a proportional-integral (PI) controller which collects and uses the passenger average waiting time in a time window of the immediate past.

\subsubsection{CVR-$\alpha$}\hfill

At each time step $k$, the number of idle vehicles is $n_{idle}(k)$. For CVR-$\alpha$,  $\lfloor n_{idle}(k) \times \alpha \rfloor$ idle vehicles with the largest $J(W_i, x_i)/J(V_i, x_i)$ value will be selected and stopped at their current positions, where $\lfloor x \rfloor$ is the greatest integer less or equal to $x$. Here we test $\alpha = 0\%, 20\%, 40\%, 60\%, 80\%, 100\%$. For $\alpha = 0\%$, one has the same setting as CVR, which never lets any idle vehicle become `not-relocated'. The case $\alpha = 100\%$ corresponds to the `Do-nothing policy', which means the empty vehicle stops at its last destination where it drops off the last passenger until it is matched with the next passenger. 

\subsubsection{CVR-PI}\hfill

The goal is to design an adaptive method to determine how many vehicles should hold their position. A PI-controller using a constant reference signal is introduced here to that end. 

The fleet size adaptation controller is operated at a constant frequency, which is chosen as once every $\Delta T'= 5$ min. At each time step $k'$, we collect and calculate the average waiting time $\bar{t}_w(k')$ and the average number of empty vehicles $\bar{n}_{idle}(k')$ in the last $\Delta T'$ minutes. The output signal can be obtained as $y(k') = \sqrt{\bar{t}_w(k') \cdot (n_{AV} - \bar{n}_{idle}(k'))} $, where $n_{AV}$ is the number of all vehicles in the AMoD system.

The error signal is defined as the difference between a constant reference and the output as 
\begin{equation}
    err(k') = y_{ref} - y(k').
\end{equation}

Then, the discrete PI controller is given by 
% \begin{equation}
%     \Delta u_{act}(k) = K_p \cdot err(k') + K_I \sum_{i=1}^k err(k') + K_D[err(k') - err(k' - 1)]
% \end{equation}
\begin{equation}
    \Delta u_{not}(k') = K_p \cdot err(k') + K_I \sum_{i=1}^k err(k'),
\end{equation}
and $u_{not}(k + 1) $ can be calculated as 
\begin{equation}
    u_{not}(k' + 1) = u_{not}(k') + \Delta u_{not}(k').
\end{equation}

In the next time step, $\lfloor u_{not}(k' + 1) \rfloor$ idle vehicles with the largest $J(W_i, x_i)/J(V_i, x_i)$ values are selected as `not-relocated idle vehicles', and the rest as `active'. More information on tuning the parameters of CVR-PI can be found in Appendix \ref{appendix}.

\subsection{Simulation Results}
\begin{figure*}[htb]
\centering
\begin{subfigure}[ht]{1\textwidth}
\centering
\includegraphics[width=1\textwidth]{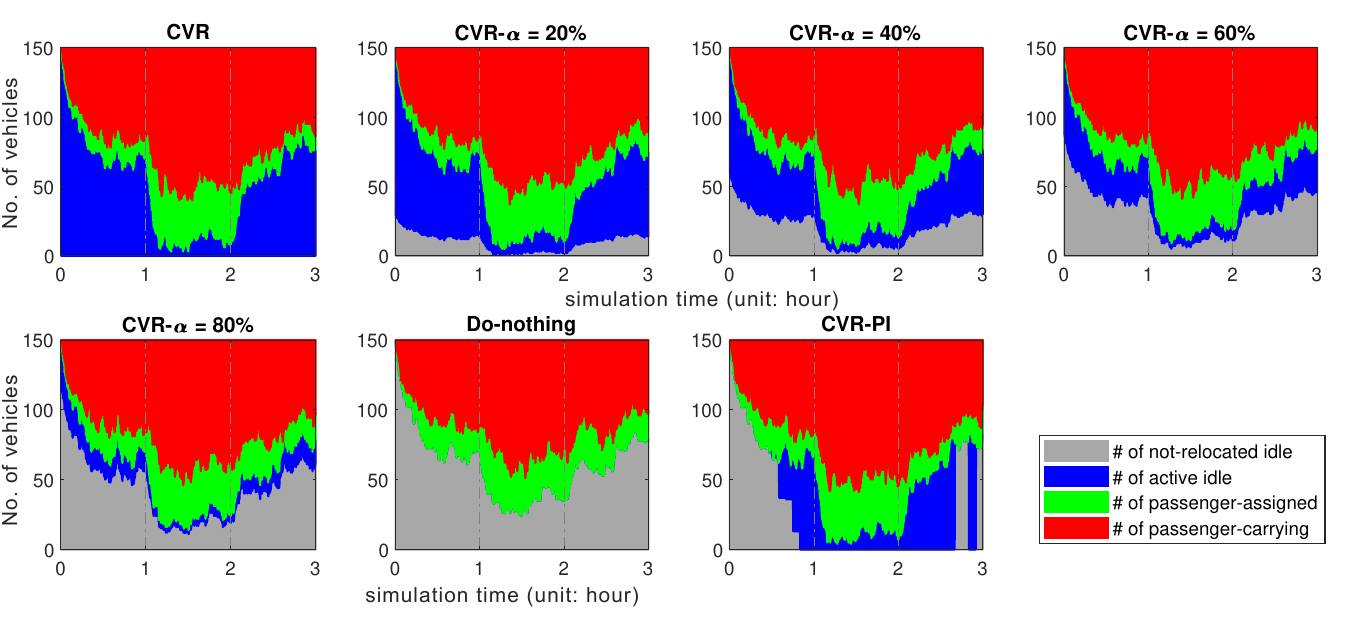}%
 \caption{No. of vehicles in different states}\label{fig:state_evo}
\end{subfigure}\\
\begin{subfigure}[ht]{0.4\textwidth}
\includegraphics[width=\textwidth]{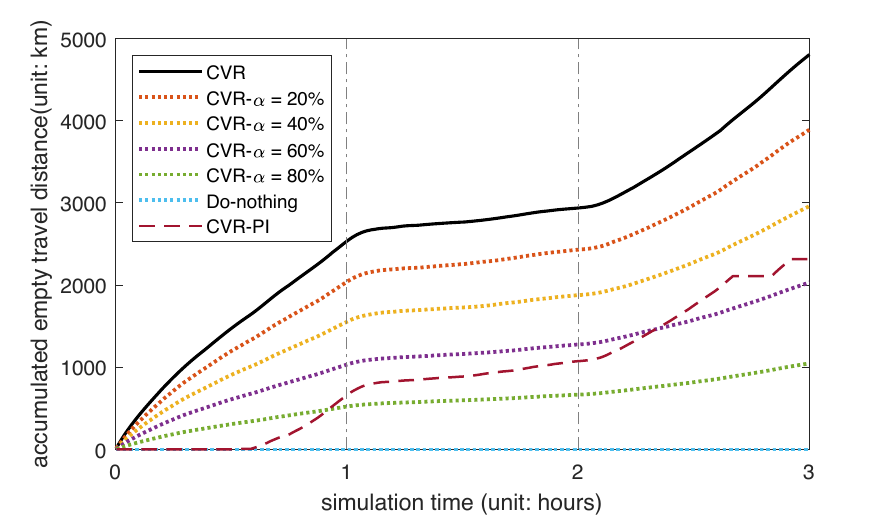}%
 \caption{Accumulated rebalancing distances}
\label{fig:accumulated_trip}
\end{subfigure}
\begin{subfigure}[ht]{0.4\textwidth}
\includegraphics[width=\textwidth]{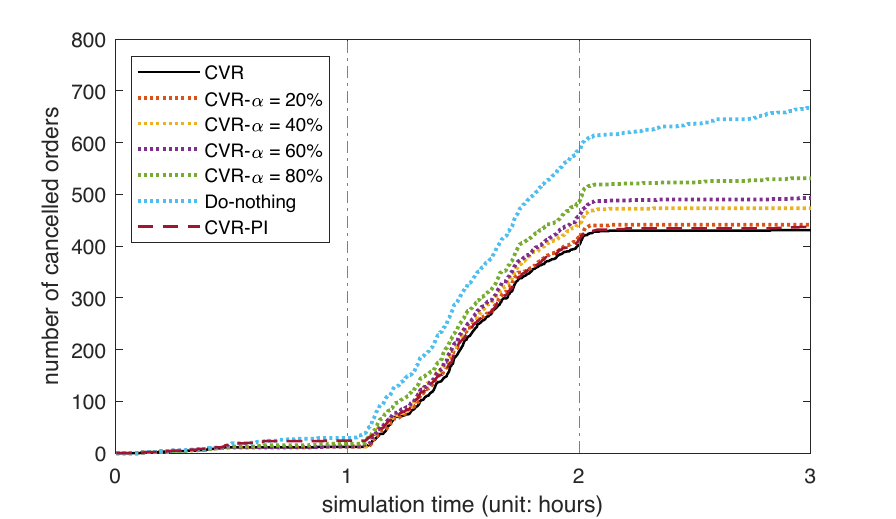}%
 \caption{Accumulated canceled orders}
\label{fig:cancel_orders}
\end{subfigure}
\caption{Comparison of CVR, CVR-$\alpha$,and CVR-PI.}
\label{fig_sim}
\end{figure*}

% As we select a subset of all idle vehicles not to move, they will hold their position, while the rest of the idle vehicles will move following the proposed coverage control algorithm. 
Here we define a performance metric for evaluating the proposed schemes: the \textit{accumulated rebalancing distance}, which is the total traveling distance caused by relocating active idle AVs. Obviously, the rebalancing distance of Do-nothing policy is always $0$ due to all empty vehicles stopping at their last destination.

Next, we present a simulation experiment over 3 hours under origin-destination trip demand imbalances $\gamma = 0.5$ with a fleet size of 150 vehicles. We compare the performances of our proposed methods with the Do-nothing policy. The performance metrics are provided in \cref{table_pi}. For CVR-PI, we have set $K_p = 0.2, K_I = 0.4$ and $y_{ref} = 60$. 

\begin{table}[ht]
  \begin{center}
    \caption{Performance of fleet size adaptation methods.}\label{table_pi}
     \vspace{2ex}
    \begin{tabular}{|p{1.8cm}||c|c|c|c|} % <-- Alignments: 1st column left, 2nd middle and 3rd right, with vertical lines in between
    \hline
       {   }&{completion} & {$\bar t_w$} & {$\bar t_{sys}$} & { accu. rebalancing}\\
       {   }&{ rate ($\%$)} & {($s$)} &{ ($s$)} &{ distance ($km$)}\\

      \hline
      CVR$ (\alpha = 0\%)$    & 82.6    &127.0  &183.2  & 4804.8    \\
      \hline
    %   CVR-PID  & 82.0    &140.7    &196.5  &2105.3\\
        %   CVR-PID  & 81.7    &136.6    &194.0  &2450.2\\
      CVR-PI  & 82.4    &139.4    & 194.2  &2314.7\\
      \hline
      CVR-$\alpha = 20\%$    &82.2    &132.5  &188.9   &3890.1    \\
      CVR-$\alpha = 40\%$    &80.9    &135.5  &195.5   &2960.0    \\
      CVR-$\alpha = 60\%$    &80.1    &137.6  &199.7   &2031.6    \\
      CVR-$\alpha = 80\%$    &78.6    &146.5  &211.5   &1047.5    \\
      \hline
      Do-nothing   &73.1    &155.1  &234.5   &0    \\
      \hline
    \end{tabular}
  \end{center}
\end{table}

It can be seen from \cref{table_pi} that CVR yields the highest completion rate and the shortest waiting time. However, since all vehicles are active and relocated all the time, the accumulated rebalancing distance is the highest. For CVR-$\alpha$, for higher values of $\alpha$, the waiting time becomes larger and the completion rate tends to decrease slightly. Generally speaking, the fleet size should be chosen properly to strike the desired trade-off between waiting time and extra traveling caused by rebalancing. It is interesting to note, however, that the CVR-PI can finish almost as many orders as CVR with the rebalancing distances reduced by $51.8\%$. Furthermore, compared to the Do-nothing policy, it can still yield significant improvements in waiting time and system time. These results indicate that the CVR-PI method is capable of achieving a good balance between system performance and rebalancing costs.

To demonstrate the simulation process more precisely, the results in \cref{fig:state_evo} depict the time evolution of the number of vehicles in various operating states (i.e., not-relocated idle, active idle, passenger-assigned, and passenger-carrying). In \cref{fig:accumulated_trip} and \cref{fig:cancel_orders}, the solid lines present the CVR method, while results of the CVR-$\alpha$ are shown by dotted lines and CVR-PI in brown dashed lines.

During the first and second hours, it can be seen from the evolution of vehicle states that the coverage control approach helps the system as the sum of blue and gray areas (i.e., the number of active idle vehicles + the number of not-relocated idle vehicles) of all algorithms are smaller than Do-nothing in \cref{fig:state_evo}. It indicates that the proposed methods are able to operate the fleet more efficiently, as a larger amount of vehicles is actively serving passengers most of the time. On the other hand, the Do-nothing policy operates with a larger amount of idle vehicles, which manifests itself as a greater amount of canceled orders, as can be seen from \cref{fig:cancel_orders}. 

\cref{fig:state_evo} shows that, at the beginning and the end of low-demand periods, CVR-PI method keeps more idle vehicles static; while during the high-demand hour, all vehicles are actively doing coverage control which guarantees completion rates as good as CVR, however with reduced rebalancing distances.

In \cref{fig:cancel_orders}, during rush hours, a steeper upward tilt to the curve for Do-nothing policy can be observed. After the high-demand period, the number of canceled orders for the Do-nothing policy still keeps increasing, while except for CVR-$\alpha = 80\%$ which produces a few more canceled orders, the other methods can manage to respond to new requests almost without any more cancellations.

\section{Conclusions}\label{conclusion}
In this paper, we proposed the application of coverage control to rebalance vehicle fleets for autonomous Mobility-on-Demand systems. Being a model-free approach, it provides a relatively simple way to give online node-level guidance for idle vehicle fleets. For compensating spatiotemporal imbalances in demand and supply, our proposed method can dynamically rebalance the spatial distribution of idle vehicles to serve more trips with reduced waiting times. To treat the non-circular areas covered by vehicles in road networks due to road geometry, we extended the original coverage control algorithm to the graph Voronoi partition. Considering the empty traveling distance due to the repositioning process, we proposed different approaches to automatically manage the active idle fleet size. Our coverage control algorithms are tested on a discrete city map using real road network geometry and trip data from the city of Shenzhen. The performances of the proposed approaches are compared with a linear programming-based rebalancing approach and a `Do-nothing' policy, and the results show clear advantages.

Simulation results depend on a pre-computed static demand density function which is estimated from historical data. However, the demand patterns are highly influenced by weather and other external effects in reality, also demand data can be skewed by measurement errors. Future research will consider using time-varying demands based on real-time trip request data for fleet rebalancing. For reference on integrating estimating density function with coverage control, see works \cite{Lee2015, Xu2020}. Additionally, it is more practical to consider a mixed system consisting of both autonomous vehicles and human drivers, with the latter potentially having inaccuracies in following the rebalancing control actions. Then besides system-wide metrics, investigating the profit of individual drivers as discussed in \cite{BEOJONE2023} is one interesting topic.
% It is of interest to investigate the influences caused by `selfish' drivers. In addition, the proposed CVR and CVR-graph algorithms are run by the vehicle itself to compute its own control actions, while one drawback
Another limitation of the current control schemes is that the fleets lack coordination with other regions in the network. One way to solve this problem is to develop a hierarchical control architecture. It can benefit from efficient coordination between the actions of upper-level controllers which operate the aggregated traffic components (for example, how many idle vehicles should travel from one sub-region to another) and those of the lower-level controllers based on the proposed CVR approaches.
% (self-management of individual vehicles at lower level, i.e., the proposed CVR methods).

\section*{Acknowledgments}
The authors would like to thank Dr. Caio Vitor Beojone for providing the discrete space simulator, and Dr. Leonardo Bellocchi for helping with preprocessing the data from Shenzhen, China. This research is supported by NCCR Automation, a National Centre of Competence in Research funded by the Swiss National Science Foundation (grant number 51NF40\_180545).

\begin{appendices}

\section{Proof of coverage control law}
\label{appendix_control}
This section provides the proof that the coverage control law (\ref{controlW}) can decrease the objective function $H(X, W)$ and asymptotically steer the agents to the centroids of their $r$-limited Voronoi cells. The proof is largely inspired by \cite{Jorge2004}.

According to the parallel axis theorem, similarly to \cite{Jorge2004}, we can rewrite the polar moment of inertia as
\begin{equation}\label{polarmomentJ}
    J(W_i, x_i) = J(W_i,C(W_i)) + M(W_i) \Vert x_i - C(W_i) \Vert^2,
\end{equation}
where $J(W_i,C(W_i))$ is the polar moment of inertia with respect to the centroid $C(W_i)$.

From \cref{coverageobjW} and \cref{polarmomentJ}, the objective function and its partial derivative with respect to $x_i$ can be expressed as:
\begin{equation}
    \begin{aligned}
    H(X,W) &= {\sum{_{i = 1}^{n}} J(W_i, x_i)} \\
    &=  \sum{_{i = 1}^{n}} {J(W_i,C(W_i))} +  \sum{_{i = 1}^{n}} M(W_i) \Vert x_i - C(W_i) \Vert^2,\\
    \dfrac{\partial{H(X,W)}}{\partial x_i} &=  2 M(W_i) ( x_i - C(W_i)).
\end{aligned}
\end{equation}

From the first-order condition $\partial{H(X,W)}/{\partial x_i} = 0$, one has that the optimal configuration involves positioning all agents at the centroids of their respective $r$-limited Voronoi cells, i.e. $x_i = C(W_i)$.
% This optimal partition of environment and critical agent positions yield the so called \textit{Centroidal Voronoi Configuration}. 

The control algorithm should minimize the coverage cost $H$ by properly allocating $n$ agents in space. To achieve this goal, the control law (\ref{controlW}) makes the position of each agent $x_i$ follow a gradient descent flow.

% By LaSalle's principle, 
% when we set the control law as \cref{controlW}, and the agent's movement obeys the dynamics in \cref{dynamics}, the objective function $H$ can be guaranteed to decrease for $\forall k_W>0$ because
For the system given by (\ref{dynamics}) and (\ref{controlW}), the objective function $H$ is guaranteed to decrease for $\forall k_W>0$ because
\begin{equation}\label{increaseH}
\begin{aligned}
\dfrac{\mathrm{d} H}{\mathrm{d} t} & =\sum{_{i = 1}^{n}} \dfrac{ \partial {H(X,W)}}{\partial x_i} \dfrac{\mathrm{d} x_i}{\mathrm{d} t}\\
%                                   & = \sum{_{i = 1}^{n}} \dfrac{\partial J(W_i, C(W_i))}{\partial x_i} \dfrac{\mathrm{d} x_i}{\mathrm{d} t}\\
                                   & = -2k_W \sum{_{i = 1}^{n}} M(W_i) \Vert x_i - C(W_i) \Vert^2 < 0.
\end{aligned}
\end{equation}

\section{PI Controller Reference Tuning}
\label{appendix}
In this section, we provide the parameter tuning process of the PI controller for the proposed CVR-PI method. With $y(k')$ reflecting the service level in the last 5 minutes, when $y(k')$ is small the fleet of active vehicles can be deemed to fulfill the current requests. In this case relocation of idle vehicles is not highly beneficial, thus when $y(k') \leq 90$, all idle vehicles will hold their current positions. Otherwise, i.e., when $y(k') > 90$, we use a PI controller to track a constant reference $y_{ref}$ to determine how many vehicles should not relocate. 
% We conduct a sensitivity analysis for tuning $y_{ref}$, the results of which are provided here.
Next, we describe the result of a sensitivity analysis that we performed for tuning $y_{ref}$.

\begin{figure}[ht]
     \centering
         \includegraphics[width=0.5\textwidth]{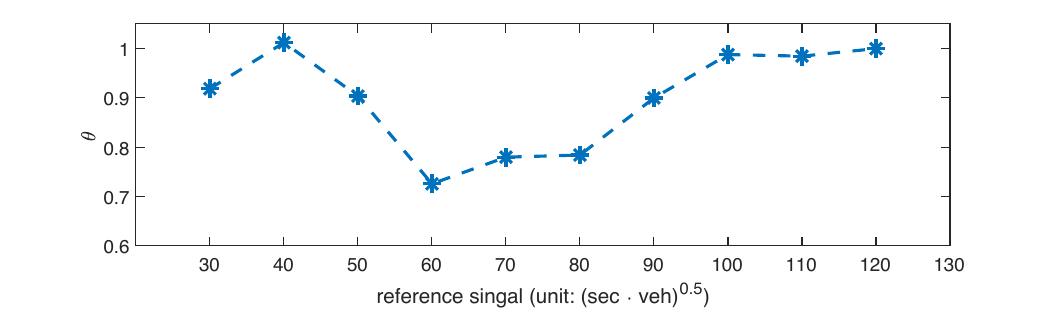}
         \caption{Tuning parameter for the reference signal}
         \label{fig:yref}
     \end{figure}

Under the same experiment settings of \cref{Experiments}, we use different values of $y_{ref}$ (from 30 to 120  with an increment of 10 ) and collect the performance metrics. The tuning criterion (denoted as $\theta$) is a normalized combination of two performance criteria, and is computed by summing up the normalized average system time (scaled to $[0, 1]$) and normalized rebalancing trips (scaled to $[0, 1]$). The results are shown in \cref{fig:yref}. 

From \cref{fig:yref}, the optimal value of $y_{ref}$ is $60$, thus this value is chosen for the simulations. Choosing $y_{ref}$ values beyond $120$ results in all vehicles being determined as active during the whole simulation period, so those values are excluded from the analysis. For the tuning of $K_P$ and $K_I$ the same procedure is repeated and the parameter values are chosen according to their resulting $\theta$ values. Note that this tuning procedure involves choosing the parameters for a specific AMoD operation scenario. More general parameter-tuning methods will be studied in future research.
\end{appendices}

\bibliographystyle{IEEEtran}
\bibliography{refers}
\begin{IEEEbiography}[{\includegraphics[width=1in,height=1.25in,clip,keepaspectratio]{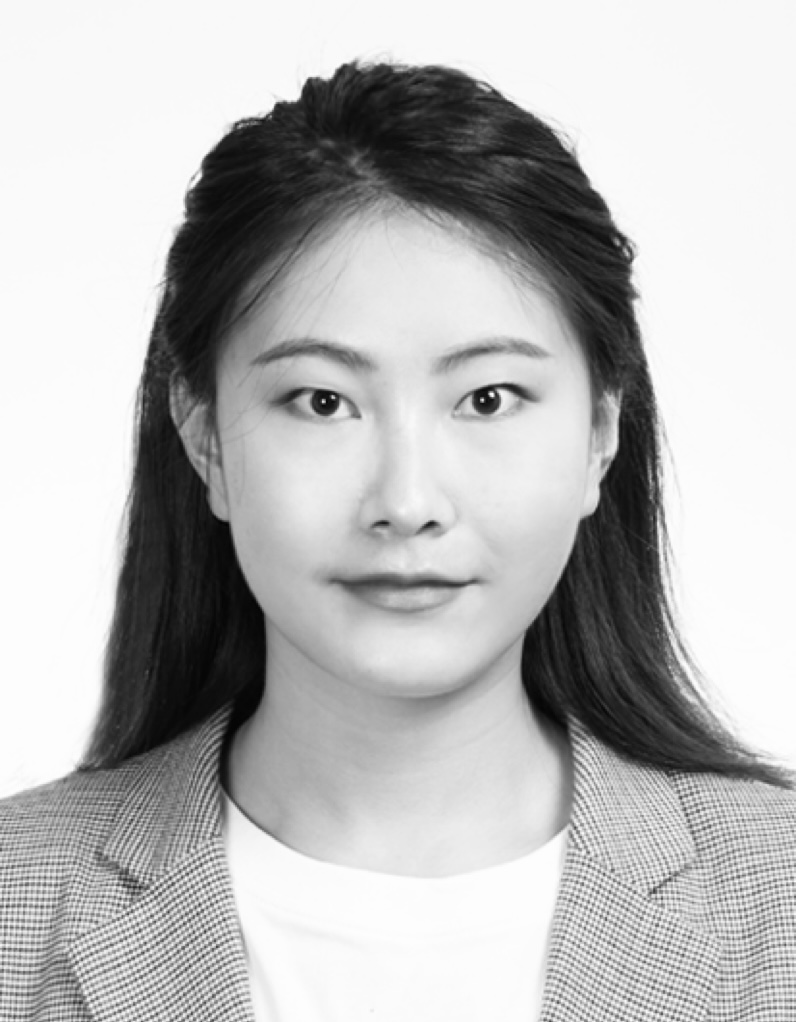}}]{Pengbo Zhu} received the B.Sc. degree in Automation in 2017 and M.Sc. degree in Control Science and Engineering in 2019 from Harbin Institute of Technology. Currently, she is a doctoral assistant at the Urban Transport Systems Laboratory (LUTS) of EPFL. Her research interests are the application of control algorithms in urban transportation systems, and the optimization of emerging mobility systems.

\end{IEEEbiography}

\begin{IEEEbiography}[{\includegraphics[width=1in,height=1.25in,clip,keepaspectratio]{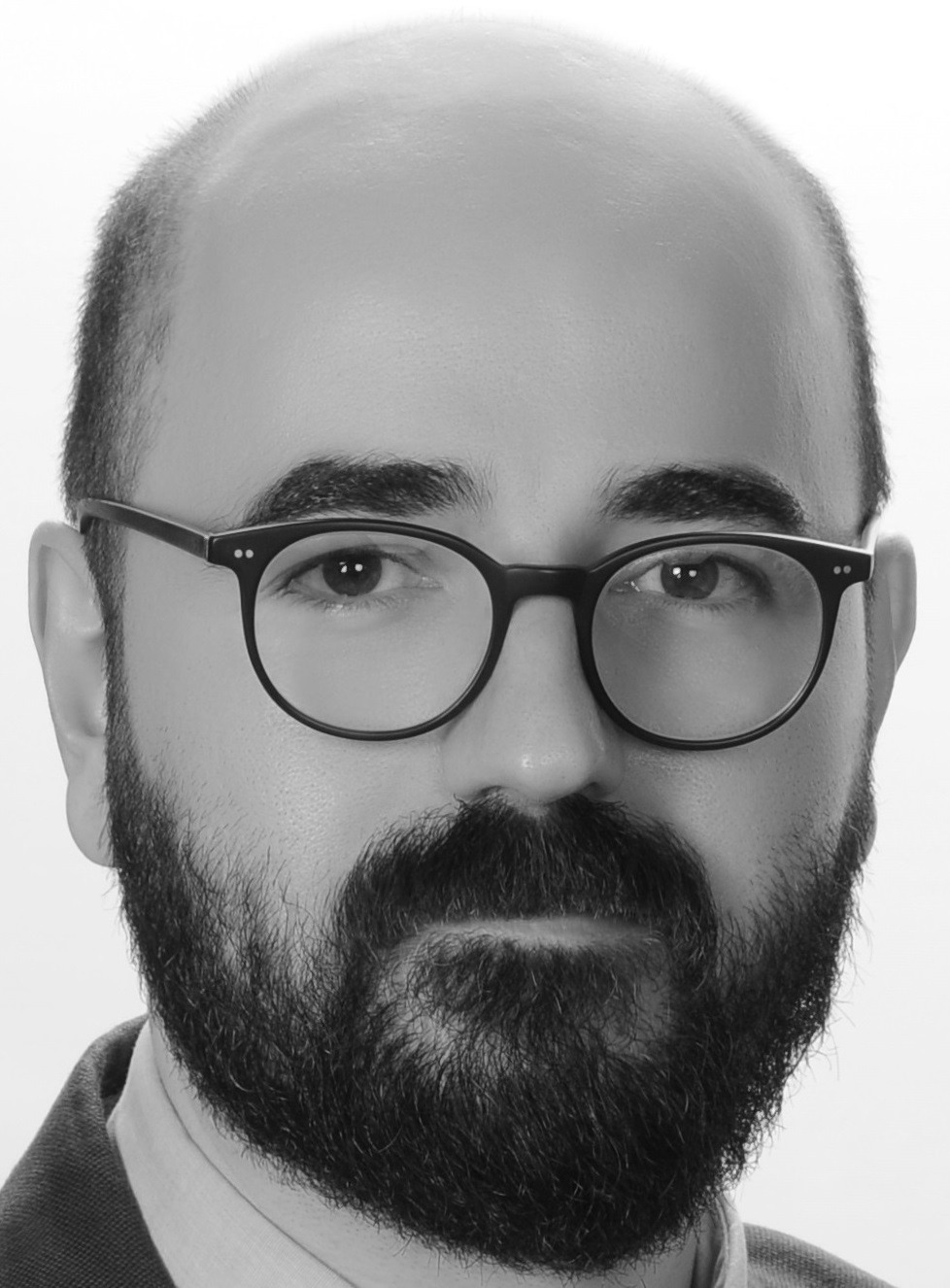}}]{Isik Ilber Sirmatel} is an assistant professor of automatic control at the Faculty of Engineering, Trakya University. He received the B.Sc. degrees in mechanical and control engineering from Istanbul Technical University, the M.Sc. degree in mechanical engineering from ETH Zurich, and the Ph.D. degree in electrical engineering from EPFL, in 2010, 2012, 2014, and 2020, respectively. Prior to joining Trakya University in 2021, he was a postdoctoral researcher at EPFL. His research focuses on applications of automatic control and optimization, with an emphasis on model predictive control of intelligent transportation systems.
\end{IEEEbiography}

\begin{IEEEbiography}[{\includegraphics[width=1in,height=1.25in,clip,keepaspectratio]{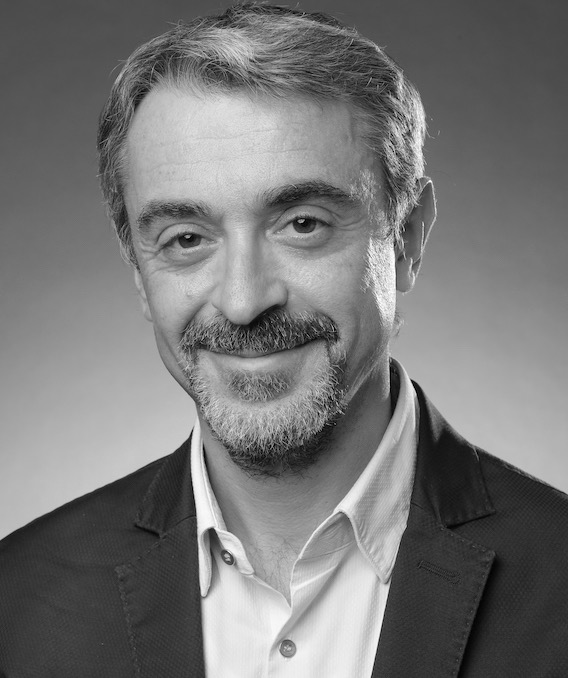}}]{Giancarlo Ferrari-Trecate}(SM’12) received the Ph.D. degree in Electronic and Computer Engineering from the Universita' Degli  Studi di Pavia in 1999. Since September 2016, he has been Professor at EPFL, Lausanne, Switzerland. In the spring of 1998, he was a Visiting Researcher at the Neural Computing Research Group, University of Birmingham, UK. In the fall of 1998, he joined the Automatic Control Laboratory, ETH, Zurich, Switzerland, as a Postdoctoral Fellow. He was appointed Oberassistent at ETH in 2000. In 2002, he joined INRIA, Rocquencourt, France, as a Research Fellow. From March to October 2005, he was a researcher at the Politecnico di Milano, Italy. From 2005 to August 2016, he was Associate Professor at the Dipartimento di Ingegneria Industriale e dell'Informazione of the Universita' degli Studi di Pavia. His research interests include scalable control, microgrids, machine learning, networked control systems, and hybrid systems. Giancarlo Ferrari Trecate is the founder of the Swiss chapter of the IEEE Control Systems Society and is currently a member of the IFAC Technical Committees on Control Design and Optimal Control. He has been serving on the editorial board of Automatica for nine years and of Nonlinear Analysis: Hybrid Systems.
\end{IEEEbiography}

\begin{IEEEbiography}[{\includegraphics[width=1in,height=1.25in,clip,keepaspectratio]{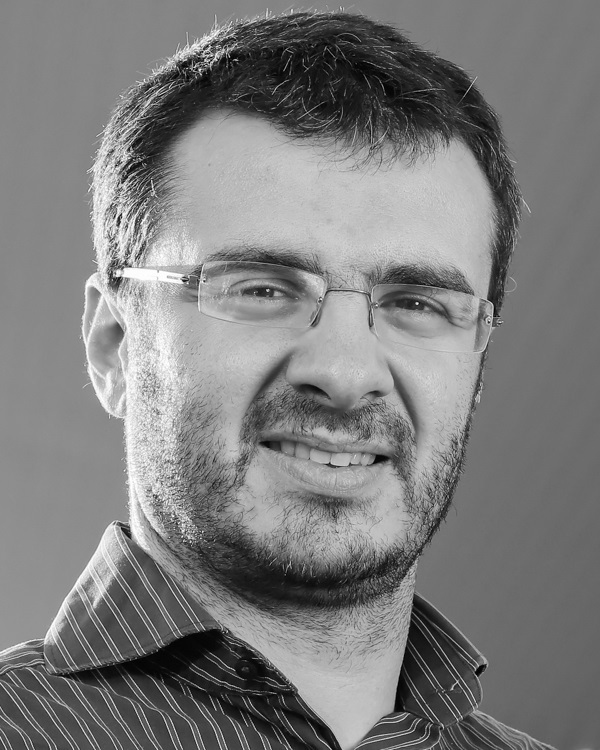}}]{Nikolas Geroliminis} is a full professor at EPFL and the head of the Urban Transport Systems Laboratory (LUTS). He has a diploma in Civil Engineering from the National Technical University of Athens (NTUA) and an M.Sc. and Ph.D. in civil engineering from University of California, Berkeley. He is a member of the Transportation Research Board's Traffic Flow Theory Committee.  He also serves as an Associate Editor in Transportation Research, part C, Transportation Science and IEEE Transactions on ITS and in the editorial board of Transportation Research, part B, Journal of ITS and of many international conferences. His research interests focus primarily on urban transportation systems, traffic flow theory and control, public transportation and logistics, on-demand transportation, optimization and large scale networks. He is a recent recipient of the ERC starting grant ``METAFERW: Modeling and controlling traffic congestion and propagation in large-scale urban multimodal networks''.
\end{IEEEbiography}

\end{document}